\documentclass[sigconf]{acmart}
\AtBeginDocument{%
  \providecommand\BibTeX{{%
    \normalfont B\kern-0.5em{\scshape i\kern-0.25em b}\kern-0.8em\TeX}}}

\copyrightyear{2023}
\acmYear{2023}
\setcopyright{rightsretained}
\acmConference[ICMI '23]{INTERNATIONAL CONFERENCE ON MULTIMODAL
INTERACTION}{October 9--13, 2023}{Paris, France}
\acmBooktitle{INTERNATIONAL CONFERENCE ON MULTIMODAL INTERACTION
(ICMI '23), October 9--13, 2023, Paris,
France}
\acmDOI{10.1145/3577190.3614166}
\acmISBN{979-8-4007-0055-2/23/10}

\usepackage{pifont}
\usepackage{multirow}
\usepackage{textgreek}
\usepackage[commandnameprefix=always]{changes} 
\usepackage{caption}
\usepackage{subcaption}
\usepackage{hyperref}

\begin{document}

\title{Implicit Search Intent Recognition using EEG and Eye Tracking:
Novel Dataset and Cross-User Prediction} 

\author{Mansi Sharma}
\affiliation{%
  \institution{DFKI, Saarland Informatics Campus}
  \city{Saarbrücken}
  % \state{Saarland}
  \country{Germany}
}\email{mansi.sharma@dfki.de}

\author{Shuang Chen}
\affiliation{%
  \institution{DFKI, Saarland Informatics Campus}
  \city{Saarbrücken}
  % \state{Saarland}
  \country{Germany}
}\email{shuang.chen@dfki.de}

\author{Philipp M\"uller}
\affiliation{%
  \institution{DFKI, Saarland Informatics Campus}
  \city{Saarbrücken}
  % \state{Saarland}
  \country{Germany}
}\email{philipp.mueller@dfki.de}

\author{Maurice Rekrut}
\affiliation{%
 \institution{DFKI, Saarland Informatics Campus}
  \city{Saarbrücken}
  % \state{Saarland}
  \country{Germany}
}\email{maurice.rekrut@dfki.de}

\author{Antonio Kr\"uger}
\affiliation{%
  \institution{DFKI, Saarland Informatics Campus}
  \city{Saarbrücken}
  % \state{Saarland}
  \country{Germany}
}\email{antonio.krueger@dfki.de}

\renewcommand{\shortauthors}{Mansi Sharma, et al.}


\begin{abstract}

For machines to effectively assist humans in challenging visual search tasks, they must differentiate whether a human is simply glancing into a scene (\textit{navigational intent}) or searching for a target object (\textit{informational intent}). Previous research proposed combining electroencephalography (EEG) and eye-tracking measurements to recognize such search intents implicitly, i.e., without explicit user input. 
However, the applicability of these approaches to real-world scenarios suffers from two key limitations.
First, previous work used fixed search times in the informational intent condition - a stark contrast to visual search, which naturally terminates when the target is found.
Second, methods incorporating EEG measurements addressed prediction scenarios that require ground truth training data from the target user, which is impractical in many use cases.
% In addition, progress on the task is limited due to the absence of publicly available training data.
We address these limitations by making the first publicly available EEG and eye-tracking dataset for navigational vs. informational intent recognition, where the user determines search times. 
%The dataset comprises $15$ subjects completing $120$ visual search tasks in an industrial workplace.
We present the first method for cross-user prediction of search intents from EEG and eye-tracking recordings and reach $84.5\%$ accuracy in leave-one-user-out evaluations - comparable to within-user prediction accuracy ($85.5\%$) but offering much greater flexibility.

% \philipp{adjust number to the ones you actually use} visual search tasks in an industrial setting. \philipp{work? some are more workbenches}
% %We have achieved an $84.5\%$ accuracy for leave-one-user-out cross-user prediction through careful feature selection, which is only slightly lower than the within-user prediction accuracy of $85.5\%$ but offers greater flexibility for various application scenarios. 

\end{abstract}

%\begin{CCSXML}
%<ccs2012>
%   <concept>
%       <concept_id>10003120.10003121.10003122.10003334</concept_id>
%       <concept_desc>Human-centered computing~User studies</concept_desc>
%       <concept_significance>500</concept_significance>
%       </concept>
%   <concept>
%       <concept_id>10010147.10010257.10010293.10010075.10010295</concept_id>
%       <concept_desc>Computing methodologies~Support vector machines</concept_desc>
%       <concept_significance>300</concept_significance>
%       </concept>
%\end{CCSXML}
%
%\ccsdesc[500]{Human-centered computing~User studies}
%\ccsdesc[300]{Computing methodologies~Support vector machines}



%%{<left> <lower> <right> <upper>}

% \begin{figure}[H]
% \includegraphics[width=\linewidth]{pic/experi (1).pdf}
% \caption{Overview of our approach to implicit classification of navigational versus informational intents.\Mansi{Use this as teaser figure}}
% \label{pipe}
% \end{figure}



\maketitle
\begin{figure}[t]
%{left bottom right top}
  \includegraphics[trim={5cm 0cm 5cm 0cm},clip,width=\linewidth]{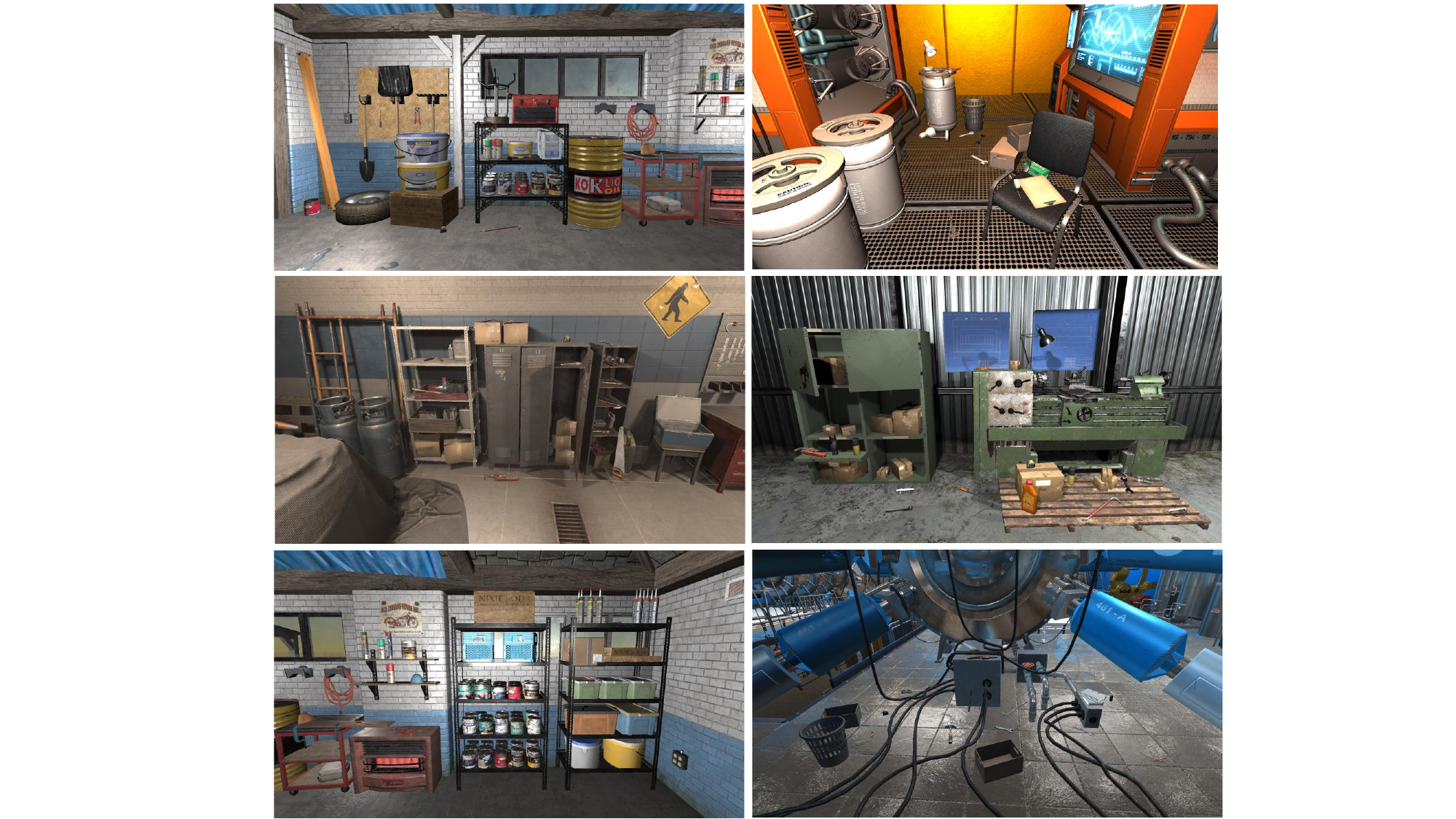}
  % \Description[Visual Search Scenes]{Examples of the visual scenes we created for our study. In total, we created 120 unique scenes.}
  \caption{Examples of the visual scenes we created for our study. In total, we created 120 unique scenes.}
  \label{scenes}
\end{figure}
% \vspace{-0.2cm}
\section{Introduction}\label{Intro}
Visual search is ubiquitous in daily life. For example, searching for a desired chocolate bar on a supermarket shelf or the wrench in a cluttered workshop.
Approaches that can automatically - and without explicit user input - infer humans' search targets have the potential to assist humans and avoid unnecessary frustration or delays~\cite{barz2020visual,sattar2015prediction}.
%Prediction of user search intents can particularly benefit Human-Machine Interaction as it can provide an implicit understanding of user intents allowing the external viewer to make predictions about the ongoing activities and pro-actively assisting in improving or building intelligent user interfaces in numerous fields like collaborative robotics in assisting their users, highlighting the significance of intention prediction in cooperative actions between humans and robots
%\cite{huang2016anticipatory,sakita2004flexible,bauer2008human}, fine-grained intent recognition for situation-aware support for mentally impaired people \cite{sonntag2015kognit}, interpreting human behavior and neurological diseases, particularly during free viewing of complicated stimuli \cite{kastrati2021eegeyenet}, to name a few. 
%Visual search is a prime example of a time-critical task that can profit from implicit intent recognition. Visual search is a task in which humans aim to identify a search target among other objects using their visual perception.
%In order to create effective assistive systems for human visual search it is crucial to 
For an assistive system to effectively help humans in visual search tasks, it has to solve one key problem before even considering which target the user might be searching for: 
the system needs to be able to recognize that the user is searching at all (i.e., having an \textit{informational intent}) and not simply looking at the scene with no such purpose in mind (called \textit{navigational intent} in the literature)~\cite{jang2014human, jang2014identification, sharma2023towards}. Previous studies on the recognition of navigational versus informational intent from EEG and eye-tracking data proved the general feasibility of this task~\cite{jang2014human, jang2014identification,park2014human,kang2015human}.
However, they suffer from limitations concerning the study design and the prediction scenarios that reduce their applicability to the real world. 
% However, study design in previous work terminate the visual search task after a fixed time (usually $5$s), without evaluating unrestricted search duration, since a typical search task is usually unbounded with respect to time. 
% The study designs terminate the visual search task after a fixed time (usually5s), unlike natural human search that is terminated once the target is found

% unlike natural human search that is terminated once the target is found.  
Using fixed time limits (usually $5$s) in visual search tasks is a common practice in previous studies \cite{kang2015human, park2014human}, as it allows for better control over the experiment and ensures that all users complete the task in a consistent amount of time. However, this approach does not allow for the evaluation of unrestricted search duration, where user has no time constraints.
%which is the time the user would take to complete the task if there were no time constraints.
This type of design allows for a more naturalistic approach to studying visual search behavior. It can provide insight into how participants prioritize search strategies and allocate their attention over time.

Concerning the prediction scenario, previous studies on multi-modal navigational vs. informational intent recognition from EEG and eye-tracking are limited to 
% within-user prediction~\cite{}\philipp{cite within user approaches if they exist, otherwise remove this part}, or 
cross-user scenarios where data from a given user can appear both in the test- and training sets
% ~\cite{kang2015human,jang2014human,jang2014identification} \philipp{check refs - or was it this one?} 
~\cite{park2014human}. 
In contrast, practical scenarios often require a model to apply to unseen target users without needing to collect training data.
% A final, but nevertheless crucial, limitation to achieve progress in search intent recognition based on EEG and eye-tracking is the lack of a publicly available dataset.
A final but crucial limitation to achieve progress in search intent recognition based on EEG and eye-tracking is the need for a publicly available dataset that will provide us valuable insights into real-world search behavior and help us generalize our intent recognition models across a larger group of users.
%The previous study with multi-modality merged all user data for classification analysis but did not analyze individual user data or any cross-user evaluation~\cite{park2014human}. \philipp{this is pretty confusing, we need to discuss how to better formulate it. Intent recognition based on EEG- and eye tracking was not yet attempted in a clean cross-user scenario, i.e. that no data from tarining users appears in test.}
%This severely limits the practical use of intent recognition approaches, as training data needs to be collected for each new user. \philipp{another limitation: no dataset publicly available}

% \philipp{wasn't there a reviewer who said previous work actually did cross-user prediction? If yes, we might want to make this a bit more precise here.}
% This severely limits the practical use of intent recognition approaches, as training data needs to be collected for each new user. \philipp{I'd focus in this paragraph on previous work and its limitations}

% \philipp{then a new paragraph to propose our solution}
Our work addresses these shortcomings by proposing a novel EEG- and eye-tracking dataset for navigational vs. informational intent classification where the search duration depends entirely on the time it takes to find the target. We intentionally select the industrial workplace scenarios where visual search often helps workers quickly and accurately identify equipment, components, and tools, which can improve efficiency, safety, and productivity.
% Our work addresses these shortcomings by proposing a novel EEG- and eye-tracking dataset for navigational versus informational intent classification in an industrial workplace scenario, where the visual search often helps workers quickly and accurately identify equipment, components, and tools, which can improve efficiency, safety, and productivity.
% Users either freely viewed a complex workplace or searched for a specific tool, where the search duration is entirely determined by the time it takes to find the target. 
% Our work addresses these shortcomings by recording a novel EEG- and eye-tracking dataset for navigational versus informational intent classification. For this dataset, we choose an industrial workplace as a highly valuable application scenario for search intent and target prediction.
%Furthermore, for the first time, we address the challenge of integrating brain signals and gaze input to infer search targets for cross-user and user-specific prediction. 
Furthermore, we propose the first multi-modal approach to the best of our knowledge for navigational vs. informational intent prediction in a strict leave-one-user-out evaluation scenario.
% \philipp{this is contradicted by related work, or at least here it sounds as multi-modality is something new, but only cross-user is actually new if I understand correctly. this has to be formulated more precisely}
% Our experiments reveal that, unlike for within-user prediction, feature selection is crucial in achieving high performance across users.
Our experiments reveal that feature selection is crucial in achieving high performance across users, unlike within-user prediction.
Our specific contributions are threefold:
\begin{enumerate}
    \item We present MindGaze\footnote{\url{https://doi.org/10.5281/zenodo.8239061}}, the first publicly available dataset for informational vs. navigational intent prediction with a large variety of different workplaces in Unity~\cite{unity} (see Figure~\ref{scenes} for examples). The dataset consists of EEG and eye-tracking recordings of $15$ participants, performing $3600$ trials.
     %\item We present the first publicly available dataset\footnote{\url{10.5281/zenodo.8239062}} for informational vs. navigational intent prediction with a large variety of different workplaces in Unity~\cite{unity} (see Figure~\ref{scenes} for examples). The dataset consists of EEG and eye-tracking recordings of $15$ participants, performing $3600$ trials.
    % \item We present the first approach for cross-user (leave-one-user-out) prediction of navigational versus informational intent, revealing that appropriate choice of features plays a crucial role in making cross-user prediction work. 
    \item We present the first approach for cross-user (leave-one-user-out) prediction of navigational vs. informational intent from EEG and eye-tracking, revealing that appropriate choice of features plays a crucial role in cross-user prediction. 
    \item We conduct extensive evaluations for within-user and cross-user scenarios and compare different multi-modal fusion strategies (early, late, and hybrid fusion). Subsequently, we perform experiments with the smaller windows of $0.5$s, $1$s, $1.5$s, and $2$s to show the potential for near real-time intent prediction applications. 
\end{enumerate}

% \noindent
% \begin{figure}[H]
% \includegraphics[trim={0cm 10cm 12cm 1cm},clip,width=\linewidth]{pic/scenes.pdf}
% \caption{Examples of the visual scenes we created for our study. In total, we created $150$ unique scenes.}
% \label{scenes}
% \end{figure}

% \begin{figure}[t]
% %{left bottom right top}
%   \includegraphics[trim={5cm 0cm 4.5cm 0cm},clip,width=\linewidth]{pic/scenes_n.pdf}
%   \caption{Examples of the visual scenes we created for our study. In total, we created 120 unique scenes.}
%   \label{scenes}
% \end{figure}
\begin{figure*}[t]
  \centering
  \includegraphics[trim={0 0 0 0},clip,width=0.47\textwidth]{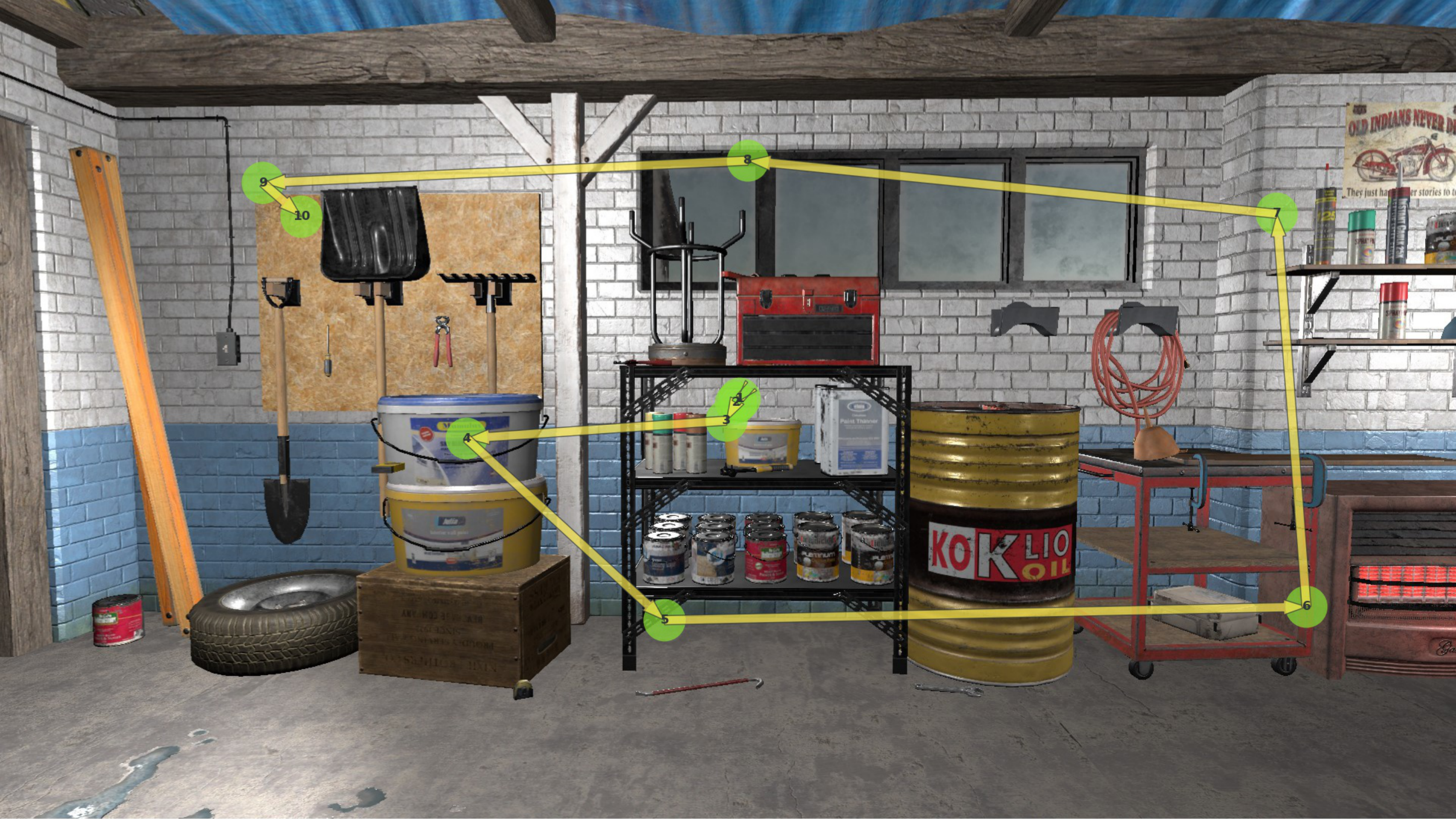}
  \centering
  \vspace{1cm}
  \includegraphics[trim={0 0 0 0},clip,width=0.47\textwidth]{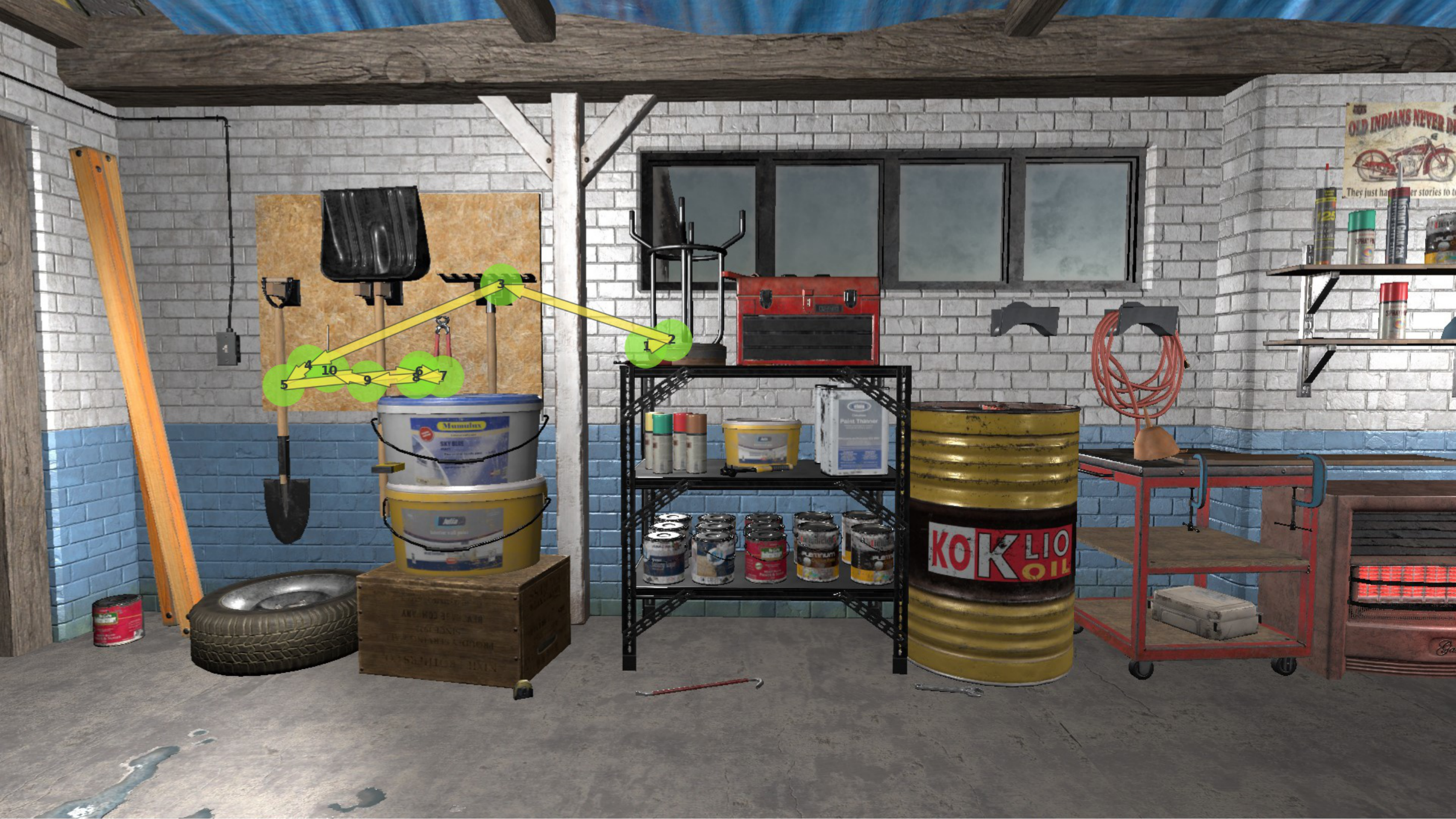}
  \vspace{-1cm}
\caption{(Left) Navigational scan path, (Right) Informational scan path, search target: \texttt{Screwdriver}}
\label{fig:scan_path}
\end{figure*}
% Proc, no short form, DOI, remove date of conference, location, Axknowledgment, authors

\section{Related Work}\label{RW}
% Humans can express their intents \philipp{``intent'' vs. ``intention'' (previous sentence). should be unified} 
%Intents refer to the thoughts one has before producing any actions~\cite{wegner2004precis}.
%Humans can express their intents either explicitly (speech, hand gestures) or implicitly (pupil analysis, neural activity)~\cite{bauer2008human,jaimes2007multimodal,ferreira2008human,jang2014human}.
The implicit (i.e., without explicit user input) prediction of user intents are of great interest in human-machine interaction (HMI), as it can help to adapt the behavior of machines without the overhead and discomfort associated with explicit input.
%Due to the lack of explicit input, 
Implicit intent recognition systems often rely on challenging-to-interpret measurements like eye-tracking, pupil dilation, or EEG~\cite{majaranta2014eye,zander2014towards}.
Previous research worked on implicitly predicting web users' click intents~\cite{slanzi2017combining}, %motor imagery intents ~\cite{zhao2020research,xu2020recognition}, 
search targets~\cite{sattar2015prediction,barz2020visual}, improving the performance of Motor Imagery task\cite{Shiweinew}
their next focus of attention~\cite{mueller20_etra,ward2016possibility,steil18_mobilehci}, or choice of ingredients~\cite{huang2015using} when preparing a meal.
%, joint action to coordinate attention and intents to achieve a common goal like carrying furniture, navigating through crowd ~\cite{sebanz2006joint,sebanz2009prediction}, building intelligent user interfaces for user's state awareness and intent prediction ~\cite{liang2019application, bednarik2012you}, to name a few.  
%
In this paper, we focus on the task of distinguishing between %(\textit{navigational intent}) 
\textrm{navigational intent} 
and \textrm{informational intent}, a pre-requisite to assist humans when searching for a target object in cluttered environments~\cite{kang2015human}.
In the following, we discuss previous work on this task.
Subsequently, we compare datasets for navigational vs. informational intent recognition recorded in previous work.

\begin{table}[t]
\caption{Datasets for search intent recognition. Availability shows whether the dataset is publicly available or not. Size is a product of the total number of users, scenes shown to each user, and intents. NA indicates missing information.}% \philipp{todo: complete sentence}}
  \label{tab:RW}
  \begin{tabular}{lccccc}
    \toprule
    Reference & Availability & Users & Modality & Size \\
    \midrule
    \citet{kang2015human} & \ding{55}  & 10  & EEG & 500 \\
    \citet{jang2014human} & \ding{55}  & 52  & Eye & 668   \\
    \citet{huang2015using} & \ding{55}  & 13  & Eye & 276   \\
    % \citet{bednarik2012you} & \ding{55}  & 12  & Eye & 8   \\
    \citet{liang2019application} & \ding{55}  & 18  & Eye & NA   \\
    \citet{jang2014identification} & \ding{55}  & 100 & Eye & 2400   \\
    \citet{park2014human} & \ding{55} & 8 & EEG + Eye & 400   \\
    \midrule
    Ours [MindGaze] & \ding{51} & 15 & EEG + Eye & 3600  \\
  \bottomrule
\end{tabular}
\end{table}

\subsection{Navigational versus Informational Intent Recognition}
%The recognition of navigational versus informational intent~\cite{park2014human, jang2014human, kang2015human} is  key prerequisite for any practical system that is supposed to assist human visual search, and that might involve the prediction of search targets in a later step

While predicting the target of visual search is an increasingly popular task in implicit intent recognition~\cite{strohm21_iccv,sattar2015prediction,barz2020visual,stauden2018visual}, few works addressed the prerequisite of any practical support system for search target prediction, namely the recognition of navigational vs. informational intent~\cite{park2014human, jang2014human, kang2015human}. 

Kang et al.~\cite{kang2015human} performed EEG-based classification of navigational and informational intents in everyday images. 
Their study setup presented an image in the navigational intent condition for $5$s, followed by the informational intent condition (search task) on the same image for a fixed search time of $5$s. 
%with a fixed search time of $5$s in the informational intent condition. 
The authors analyzed the differences in phase-locking value (PLV) to classify intents in a within-user prediction scenario. 
% \added{in a within-user prediction scenario}. \philipp{correct?} Yes
The results showed severe over-fitting with a significant gap between the train (> $99\%$) and test accuracy (~$57.1\%$ to $77.4\%$). 

Using the same concept of sequential navigational and informational intent tasks, Jang et al.~\cite{jang2014human} classified human implicit navigational and informational intent based on the eyeball movement pattern and pupil size variation characteristics in a visual search task. For evaluation, authors used $25$ users for training and other $27$ users for testing, reaching a mean accuracy of $85.26\%$ with an SVM classifier.
In a follow-up work~\cite{jang2014identification}, the authors performed hierarchical classification to further differentiate states in task-oriented visual searches, such as intent generation, intent maintenance, and intent disappearance. 
%\philipp{what does this mean?}
Authors collected data from $100$ users and used $40$ random samples for training and $60$ for testing, reaching $90.36\%$ with an SVM classifier. 
In both studies~\cite{jang2014human,jang2014identification}, users had to search for different numbers of objects in indoor- and outdoor scenes, e.g., the cup and bottle in an indoor image or all humans in an outdoor image. 
%with a pre-defined area-of-interest (AOI) in advance. 
%Consequently, their results do not directly apply to scenarios where users search for a specific target object. 
% \philipp{this statement falls out of the blue - I think it refers to a sentence that appears quite a bit earlier} 
%\Mansi{I think we should reformulate this and it's better if we do not make this strong claim probably and try to argue with other reasonable points}
%\Mansi{As a drawback, I think it is better if we state the number of distinct stimuli,  6 and 8 different images, which are very less as compared to 120 }
Although \cite{jang2014human,jang2014identification} did 
% \philipp{``did'' - use simple past mostly in related work} 
not explicitly mention fixed search times, they used 
% \philipp{same here} 
the same experimental setup as~\cite{kang2015human}, which has a fixed amount of time ($5$s) to perform the search task.

In a follow-up work to~\cite{jang2014human}, Park et al.~\cite{park2014human} proposed a multi-modal approach combining EEG and eye-tracking features while following the same experimental design principle as in~\cite{kang2015human}, i.e., using $5$s for both navigational and informational intent conditions.
The authors neither followed a pure within-user nor a pure cross-user evaluation approach.
They trained their model on several users, but samples from the same user could appear both in training- and the test set.
With an early fusion approach, they
%The authors used early fusion by concatenating EEG and eye-tracking features and 
improved classification accuracy by $5\%$ over uni-modal baselines.
While these results indicate the utility of a multi-modal approach,
%joining EEG with eye-tracking features, 
a comparison between different fusion methods (early, late, and hybrid fusion) was not presented.
Furthermore, the applicability of their approach is limited by the fixed search times and by not employing a strict cross-user prediction scenario, i.e., where data from a single user can only be either in the train- or in the test set.
%Furthermore, their study remains limited by the artificial search tasks and fixed search times of $5$s as in~\cite{kang2015human}.
%Crucially, previous intent prediction methods did not evaluate the performance when restricting the input sequence to smaller time windows, making it challenging for real-time interpretation. 
% \philipp{most of our evaluations also do this. better phrase it like this ``they do not evaluate performance when restricting the input sequence to...''}

%\philipp{TODO: incorporate the information about the evaluation scenario of each approach (within- user, cross-user, or mixed (explain)}
%\Mansi{within-user: Kang et al. (EEG) , mixed is Park et al.(EEG + eye), Jang et al both papers, kind of cross-user, it is in the text, they are completely different train and test group basically
%}

In contrast to previous work, search times in our study are entirely determined by the time it takes participants to find the target.
Furthermore, we, for the first time, study multi-modal prediction of navigational vs. informational intents in a strict cross-user evaluation scenario.
%Furthermore, we instructed users to search for a specific target object (e.g., a hammer), which makes our study setup applicable to practical scenarios for visual search assistants.  

%Finally, we investigate different approaches to fuse modalities (early, late, hybrid) and evaluate our implementation pipeline not only on the entire sequence of data but with varying window sizes of search duration to show the potential towards real-time intent prediction. 
% \vspace{-0.8cm}

\subsection{Datasets for Navigational versus Informational Intent Recognition}
% We provide an overview over the datasets used by previous works on navigational versus informational intent recognition in \autoref{tab:RW}. 
% Most datasets consist of eye-tracking recordings exclusively~\cite{jang2014identification,liang2019application,huang2015using,jang2014human}, only two datasets involving EEG recordings were presented in previous work~\cite{park2014human,kang2015human}, one of them also containing eye-tracking recordings~\cite{park2014human}.
% The number of users in eye-tracking datasets is usually larger than the number of users in datasets with EEG recordings (8-10 users), likely a result of the time-consuming procedure required to set up EEG recordings.
% The number of trials varies significantly across previously recorded datasets, ranging from $276$ to $2400$ trials.
% With $3600$ trials 
% % \philipp{check consistency with table}
% , our novel dataset of EEG and eye-tracking recordings has a larger number of trials than any previous dataset on informational versus navigational recognition. 
% We recorded $15$ users, making it also largest dataset for navigational versus informational intent recognition with EEG recordings in terms of number of users.
% Most importantly, none of the previously recorded datasets for navigational versus informational intent recognition is publicly available, severely limiting progress on this task.
% To address this shortcoming, we will make our full dataset publicly available upon publication.

We provide an overview of the datasets used by previous works on navigational vs. informational intent recognition in \autoref{tab:RW}. 
Most datasets consist of eye-tracking recordings exclusively~\cite{jang2014identification,liang2019application,huang2015using,jang2014human}, only two datasets involving EEG recordings were presented in previous work~\cite{park2014human,kang2015human}, one of them also containing eye-tracking recordings~\cite{park2014human}.
The number of users in eye-tracking datasets is usually higher than those in datasets with EEG recordings ($8$-$10$ users), likely due to the time-consuming procedure required to set up EEG recordings.
The number of trials varies significantly across previously recorded datasets, ranging from $276$ to $2400$ trials.
With $3600$ trials 
% \philipp{check consistency with table}
our novel EEG and eye-tracking recordings dataset has a much higher number of trials on informational vs. navigational recognition than any previous dataset. 
We recorded $15$ users, making it the largest dataset for navigational vs. informational intent recognition with EEG recordings regarding the number of users.
Most importantly, none of the previously recorded datasets for navigational vs. informational intent recognition is publicly available, severely limiting progress on this task.
\begin{table*}[t]
  \caption{List of extracted features with PyEEG and added statistical features}
  \vspace{-0.3cm}
  \label{PyEEG-features}
  \centering
  \setlength{\tabcolsep}{40pt} % Default value: 6pt
\renewcommand{\arraystretch}{1} % Default value: 1
  \scalebox{1}{
  \begin{tabular}{lll}
    \toprule
    %\multicolumn{2}{c}{Part}                   \\
    %\cmidrule(r){1-2}
    \textbf{Feature name}     & \textbf{Description}     \\
    \midrule
    Power spectral intensity & distribution of signal power over frequency \vspace{-0.1cm} \\ &  bands: delta, theta, alpha, beta, and gamma     \\
    \hline
    Petrosian Fractal Dimension   & ratio of number of self-similar pieces versus 
    \vspace{-0.1cm}  \\ & magnification factor      \\ \hline
    Hjorth mobility and complexity     &  mobility represents
    the proportion of the 
    \vspace{-0.1cm}  \\ & standard deviation of the power spectrum
    \vspace{-0.1cm}  \\ &  Complexity  represents the change in frequency       \\ \hline
    Higuchi Fractal Dimension     & computes fractal dimension of a time \vspace{-0.1cm}  \\ & series directly in the time domain        \\
    \hline
    Detrended Fluctuation Analysis & designed to investigate the long-range 
    \vspace{-0.1cm} \\ & correlation in non-stationary series      \\ 
    \hline
    Skewness & measure of asymmetry of an EEG signal  \\
    \hline
    Kurtosis & used to determine if the EEG data has peaked 
    \vspace{-0.1cm}  \\ & or flat with respect to the normal distribution
    \\
    \hline
    Minimum, Maximum, and Standard deviation   &  measure of variability of an EEG signal  \\
    \bottomrule
  \end{tabular}
  }
\end{table*}
% %{left bottom right top}

% \vspace{-0.2cm}
\section{Dataset}\label{Data}
In the following, we describe the data recording and descriptive statistics on the dataset. We provide detailed statistics at the dataset website (see ~\autoref{Intro}).

% \vspace{-0.2cm}
\subsection{Data Recording}

\paragraph{Participants} We recruited $15$ volunteers ($5$ female and $10$ male) aged between
$20$ and $35$ years old (µ = $27.46$ , ~σ~=~$4.22$~). All participants had normal or corrected to normal vision, and none were exposed to the study design before. The study was approved by our institution's Ethics and Hygiene Board\footnote{https://erb.cs.uni-saarland.de/}. 

% Users not associated with our institution obtained suitable compensation for participating in the experiment. 
% The study was approved by the ethical review board of the Faculty of Mathematics and Computer Science at Saarland University\footnote{https://erb.cs.uni-saarland.de/}.

\begin{figure}
%{<left> <lower> <right> <upper>}
\includegraphics[trim={0cm 5.2cm 1cm 1cm},clip,width=\linewidth]{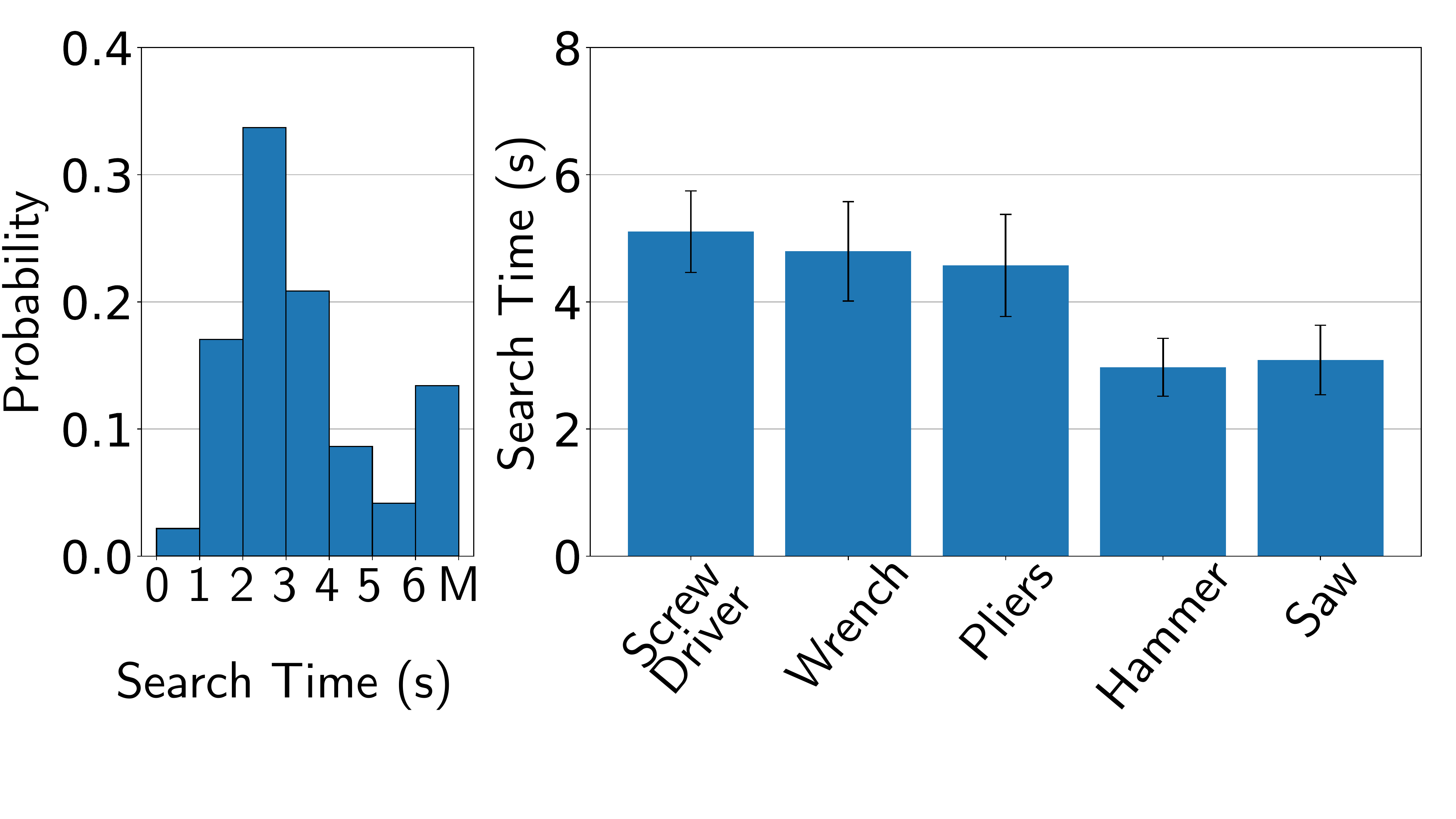}
\caption{(Left) Distribution of search times for all users. All search times larger than $6$s are accumulated in the rightmost bar, where M is the max duration. (Right) Search durations for individual tools.} 
% \philipp{font in plot too small}}
\label{fig:search_time}
\end{figure}
%of the faculty of mathematics and computer science.
% \begin{figure}[H]
% %{<left> <lower> <right> <upper>}
% \includegraphics[width=\linewidth]{pic/ex.pdf}
% \caption{Overview over one trial of our experiment, including navigational and informational intent conditions.}
% \label{exp_protoo}
% \end{figure}

% \begin{figure}
% \centering
% \begin{subfigure}{\textwidth}
%   \centering
%   \includegraphics[scale=0.5,trim={1cm 0cm 1cm 0cm},clip,scale=0.5]{pic/Searchn_resized.pdf}
%   \caption{}
%   \label{fig:search_time(a)}
% \end{subfigure}%
% \begin{subfigure}{\textwidth}
%   \centering
%   \includegraphics[scale=0.7,trim={1cm 0cm 1.5cm 1.5cm},clip,scale=0.7]{pic/search time for each tool new.pdf}
%   \caption{}
%   \label{fig:search_time(b)}
% \end{subfigure}
% \caption{(a) Distribution of search times for all users. All search times larger than 6 seconds are accumulated in the rightmost bar. (b) Search durations for individual tools}
% % \label{fig:search_time(b)}
% \end{figure}

\paragraph{Hardware Setup} 
% \philipp{always use simple past when describing what you did in your study (I have already unified it - just a note for the future)}
To display visual stimuli, we used a monitor with a resolution of $1920$ x $1080$ and screen brightness of $300$ cd/m2. 
We used the LiveAmp $64$ channel system\footnote{https://brainvision.com/products/liveamp-64/
} by Brain Products to record the EEG signals with a sampling frequency of $500$~Hz. Electrodes were placed according to the $10$-$20$ international electrode placement system \cite{electrodes}. 
% This configuration effectively captures the spatial information in brain responses by covering the whole scalp of the participant \cite{electrode}. 
To record eye-tracking data, we used wireless Tobii pro fusion\footnote{https://www.tobii.com/products/eye-trackers/screen-based/tobii-pro-fusion}  attached to the monitor's lower bezel with a sampling frequency of $250$ Hz.
The device was calibrated at the start of the experiment for each participant, using two coordinate systems. One is a $2$D system that spans the monitor with $(0,0)$ in the top right corner of the experiment setup monitor screen and $(1,1)$ in the bottom left. The second is a $3$D coordinate system for the experiment room, which measures the distance from the eye to the eye-tracker.
% We recorded gaze data with a sampling frequency of $250$ Hz. 
% and is needed mainly for I-VT filter computations.
EEG and eye-tracking data were synchronized via particular time-locked events, for example, the start and end of the navigational and informational stimulus. 

\paragraph{Stimuli} 
We designed $120$ realistic industrial scenes in Unity to simulate industrial environments, such as assembly units, manufacturing and production facilities, industrial labs, garage and repair workshop, and many more (see \autoref{scenes} for examples). While crafting the scenes, we portrayed different clutter layers through the chaotic arrangement of parts of machines, tools, workbench, and others, as evident in \autoref{scenes}. We incorporated various tool locations (inside a cupboard, on the floor, and in some unexpected areas) and orientations in the industrial scenes to create different levels of complexity. We used five tools - Hammer, Pliers, Saw, Screwdriver, and Wrench - as our target stimulus.
% \philipp{number of scenes + what were your goals in designing these scenes (e.g. high variability, different levels of clutter), and how did you make sure to achieve them?} 
% \philipp{what does ``surprise'' mean here?} 

%\subsection{Study procedure}\label{pr}
% \philipp{justify the study design (why do you present the scene before they start searching?) - also use the previous work}

% Intereference, work style, different ways of sorting and tool arranging, work-habits, car repair garage, different experience
\paragraph{Procedure}\label{pr}
% \philipp{check whether label pr still makes sense}
Participants were given a general introduction to the study, where we explained the experiment's purpose and procedure and discussed the anonymity and privacy of their collected data. Next, we gathered participants' consent and asked them to complete a demographic questionnaire. 
Participants were seated in a comfortable chair such that the distance between the user and the screen was $60$cm.
We mounted the EEG cap on their head, filling the electrodes with gel. Following common practice~\cite{imp} for noise reduction, we kept electrode impedances below $25$ k$\Omega$ throughout the experiment. Overall, the preparation time was about $30$ min. Our experimental design followed previous research \cite{park2014human,jang2014human,jang2014identification}, which presents the scene first in the navigational intent condition, followed by the search task (informational intent condition). This approach mirrors a typical workplace scenario where individuals often start searching for items within a scene that is already familiar to them.
In particular, a single recording of the experiment consisted of $3$ steps.
%as shown in \autoref{pipe}. 
In Step~1, we presented the scene for $5$s. The participant glanced over the scene without knowledge of the target tool. 
In Step~2, we show the target tool for $5$s. In the final Step~3, the participant searches for the displayed target tool in the scene until the tool is found. 
Once the participant fixated on the target tool for more than $1$s, a red highlight appeared around the tool, which changed to green in the case of prolonged fixation of $1$s
%\philipp{after what time?}
, indicating that the tool was successfully located. 
%As feedback to the participant that they found the correct tool, a red highlight appears around the periphery of the tool when fixated, which later changes to a green color to verify that the object is appropriately fixated. To ensure the correctness of this mechanism and avoid false highlights, we keep a buffer time of $1$s for the highlight to appear. 
%We took equal duration for Navigational and Informational Intent within each sample as our feature extraction module expects the input to have the exact dimensions and to remove any bias for the classifier due to varying time durations. 
The $120$ unique scenes were split randomly into four equal sessions, with breaks in-between sessions. The sequence of the industrial scenes and the target objects are randomized with an average experiment duration of $90$ min per participant.
% The sequence of industrial scenes and the sequence of target objects are randomized. On average, the experiment lasted $90$ min per participant.

%Combined Figure
% \begin{figure}[t]
% %{<left> <lower> <right> <upper>}
% \includegraphics[trim={0cm 9cm 0cm 7cm},clip,width=\linewidth]{pic/merged.pdf}
% \caption{(Left) Distribution of search times for all users. All search times larger than 6s are accumulated in the rightmost bar. (Right) Search durations for individual tools \philipp{font in plot too small}}
% \label{fig:search_time}
% \end{figure}

% \begin{figure}[H]
%   \centering
%   \includegraphics[trim={0 0 0 0},clip,width=0.49\linewidth]{pic/histogram_inf_durations.pdf}
%   \centering
%   \includegraphics[trim={0 0 0 0},clip,width=0.49\linewidth]{pic/search_time_for_each_tool_wb4.pdf}
% \caption{(Left) Distribution of search times for all users. All search times larger than 6 seconds are accumulated in the rightmost bar. (Right) Search durations for individual tools \Mansi{Todo: Figure x n y ticks}}
% % \label{fig:search_time}
% \end{figure}

% \vspace{-0.cm}
\subsection{Descriptive Statistics} \label{Analysis}
% \philipp{we need to explain the missing data (i.e. excluding one session at the end somewhere.) here would be a good point. there can also be other ways of communicating it to make it sound a bit more professional, e.g. we can already say in the procedure that we recorded 3 session (and make a footnote about the 4th that was excluded). In any case numbers need to match up with the dataset comparison table. We also need to discuss data we removed due to synchronization issues.}
% We collected the data from $15$ users by showing them $120$ scenes. However, we removed a few samples with our preliminary data analysis because of the synchronization issue between EEG and Eye-tracking. Despite this setback, we could still use an average of $102$ samples per user for our results and experimental evaluation.
% We collected the data from $15$ users by showing them $120$ scenes. 
% % However, we removed a few samples (on an average of $18$ scenes per user) with our preliminary data analysis because of the technical issue with EEG and Eye-tracking data recording. 

We collected the data from $15$ users who viewed $120$ scenes each. The data synchronization is performed based on the available timestamps.
We pre-processed the collected dataset and filtered out scenes (an average of $18$ scenes per user) due to technical issues with EEG and Eye-tracking data recording. Figure~\ref{fig:scan_path} showed the scan paths of an example participant. % for navigational and informational conditions. 
In the navigational condition, \autoref{fig:scan_path} (Left), the users' attention is spread across the scene. In \autoref{fig:scan_path} (Right), the user performs a more focused search for finding the target, which is a \texttt{Screwdriver}. Unlike existing studies~\cite{kang2015human,park2014human,jang2014human}, the search times in our informational intent condition were entirely determined by the time participants took to find the target. To illustrate the importance of this choice, Figure \ref{fig:search_time} (left) visualizes the distribution of search times. % collected in our experiment.
The search time follows a left-skewed Gaussian distribution with a peak between $2$s and $3$s. 
While the probability diminishes continuously for larger search times. 
%, we also did observe long search times beyond $10$s.
Overall, it is evident that the variation in search times is large, supporting our approach of not setting a fixed search time in contrast to previous work~\cite{kang2015human}. Figure~\ref{fig:search_time} (right) describes the search time with respect to five different target tools for all users. The search time appears to be connected with the size of the tool. For example, \texttt{Screwdriver}, \texttt{Wrench}, and \texttt{Pliers} have comparable sizes and hence, comparable search times. %As \texttt{Hammer} and \texttt{Saw} are bigger than other tools, they have shorter search times. 

%with the longest search time being $55$ sec.
%Figure~\ref{search_time}(a) accumulates all probabilities between $5$ and $55$ into a single bin due to space constraints. 

%We created the scan path by using \cite{dalmaijer2014pygaze}.  
%\philipp{it would be interesting to compare simple features like saccade length or fixation durations between the two conditions. or did we say this will be done in results section?}\Mansi{I am wondering about this. If we only show Eye features not EEG, would this be a problem?}

\begin{figure*}
  \includegraphics[trim={1.5cm 3cm 4.2cm 5.3cm},clip,width=\textwidth]{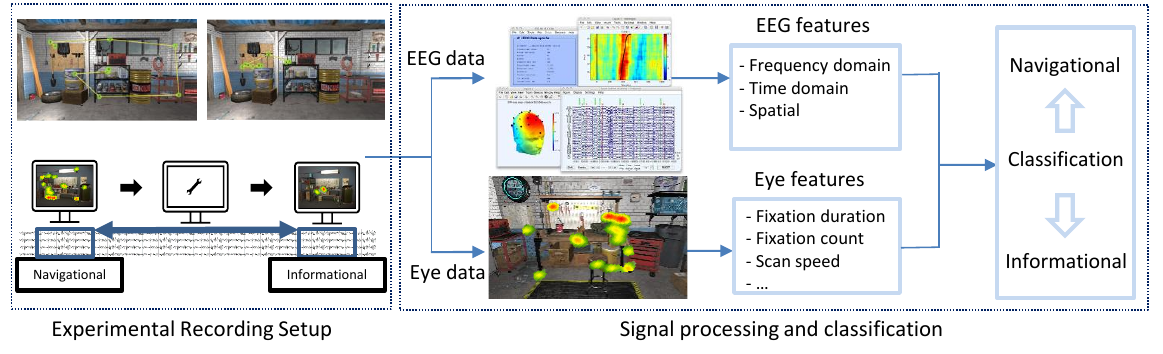}
  \caption{Overview of our approach to implicit classification of navigational versus informational intents.}
  %Each recording setup consists of two trials, one for each intent.}
  % \philipp{according to our notion of trial in the dataset comparison table, these would be two trials - one navigational and informational intent trial.}}
  % \philipp{in the image it says ``free-viewing'' versus ``searching''. I think we should rather use navigational versus informational intent to align with our main text and previous work.}}
  %\Mansi{structure teaser figure}}
  \label{pipe}
\end{figure*}

% \vspace{-0.2cm}
\section{Methods}\label{Methods}
% \philipp{I can iterate this already}
%We thoroughly investigated and determined the most effective approach for our novel dataset and challenging cross-user prediction. Our analysis includes pre-processing, feature extraction, fusion mechanisms, and three classification algorithms. Our findings demonstrate that the early fusion of EEG and eye tracking features is the most effective approach, as depicted in Figure~\ref{pipe}. 
In the following, we detail pre-processing and feature extraction pipelines for EEG and eye-tracking data and present our classification and multi-modal fusion approaches. Figure~\ref{pipe} showed an overview of data collection, signal processing, and classification.

% \vspace{-0.2cm}

\subsection{Pre-processing}

% We removed the end of the informational condition during which the eye only rests on the target object to highlight the color to know if the correct tool was located (see Section~\ref{pr}).
%duration in the Informational part where the eyes are resting because the participant is only fixating on the tool successfully located, see section~\ref{pr}. 

\paragraph{Eye data pre-processing} As humans extract visual information during fixations, visual search tasks are commonly analyzed based on fixations~\cite{barz2020visual, stauden2018visual}.
We used the I-VT filter, a velocity-threshold fixation detection approach~\cite{olsen2012tobii}. 
Overall, we follow the seven-step approach taken by~\cite{trabulsi2021optimizing}:
\textit{Gap fill-in} helps replace missing samples, which might occur due to unforeseen disturbances causing short gaps in the data. 
\textit{Eye selection} averages the position data from the left and the right eye. We applied \textit{Noise reduction} to smooth out the noise while preserving the features based on the moving median method. We then used the \textit{Velocity calculator} to associate each gaze sample with a velocity. 
To classify the sample as either a part of fixation or not, we applied the \textit{I-VT classifier}. Subsequently, we employed \textit{Merge adjacent fixations} to correct erroneously split fixations due to noise. 
Lastly, using \textit{Discard short fixations}, we discard fixations which are too short to be relevant in visual search.  
%https://www.ncbi.nlm.nih.gov/pmc/articles/PMC8606821/

\vspace{-0.2cm}
\paragraph{EEG data pre-processing} We applied high-pass filters at a cutoff frequency of $1$~Hz to clean the EEG data, followed by a notch filter between $48$~Hz and $52$~Hz~\cite{high} and a low-pass filter at a cutoff frequency of $40$~Hz~\cite{klug2021identifying}. 
Then bad channels were removed and interpolated using signals in good locations based on
the spherical interpolation method. 
In the next step, we referenced
channels to a common average reference~\cite{ludwig2009using}. 
To reduce the correlation between electrodes, we executed independent component analysis using the second Order Blind Identification (SOBI) algorithm~\cite{delorme2004eeglab}, followed by subsequent automated IC\_Label rejection (muscle, heart, and eye components with a $95$\% threshold). Lastly, we extracted specific time windows from the continuous EEG signal~\cite{levy1987effect}, with reference to the stimulus onset in the pre-processed data. 
We took equal duration for Navigational and Informational Intent within each sample, a prerequisite of one of the feature extraction methods (Common Spatial Pattern). We followed the same approach for other feature extraction methods to ensure a fair comparison.
\begin{table*}
\caption{Overview of eye movement features based on fixation events, saccadic eye movements, and the scanned area}
\vspace{-0.3cm}
  \label{eye-features}
  \centering
   \setlength{\tabcolsep}{12pt}
\begin{tabular}{lll} 
\hline
& \textbf{Feature} & \textbf{Description} \\
\hline
\multirow{4}{4em}{fixation-based} & fix\_n & Number of fixations \\ 
& fixn\_dur\_sum & Sum of fixation durations \\ 
& fixn\_dur\_avg  & Mean of fixation durations \\
& fixn\_dur\_sd  & Standard deviation of fixation durations  \\ 
\hline
\multirow{7}{4em}{saccade-based} & scan\_dist\_h  & Sum of horizontal amplitudes of all saccades, normalized by a factor w, i.e., screen width\\ 
& scan\_dist\_v & Sum of vertical amplitudes of all saccades, normalized by a factor h, i.e., screen height \\ 
& scan\_dist\_euclid  & Sum of Euclidean distances of normalized amplitudes of all saccade \\
& scan\_hv\_ratio   & Ratio of horizontal to vertical amplitudes: scan\_dist\_ h$/$scan\_dist\_v  \\ 
& avg\_sacc\_length  & Average saccade amplitude: scan\_dist\_euclid$/$(fix\_n - 1)  \\ 
& scan\_speed\_h    & Horizontal saccade velocity: scan\_dist\_h$/$scan\_time  \\ 
& scan\_speed\_v   & Vertical saccade velocity: scan\_dist\_v$/$scan\_time \\ 
& scan\_speed   & Saccade velocity: scan\_dist\_euclid$/$scan\_time \\ 
\hline
\multirow{5}{4em}{area-based} & box\_area  & Area spanned by summed saccade amplitudes: scan\_dist\_h $\times$ scan\_dist\_v\\ 
& box\_area\_per\_time  & The box\_area normalized by the scan time: box\_area$/$scan\_time \\ 
& fixns\_per\_box\_area  & Number of fixations per scanned area: fixn\_n/box\_area \\
& hull\_area\_per\_time   & The hull area normalized by the scan time: hull\_area$/$scan\_time \\ 
& fixns\_per\_hull\_area  & Number of fixations per convex hull area: fixn\_n/hull\_area  \\ 
\hline
\end{tabular}
\end{table*}

% \vspace{-0.1cm}
\subsection{Feature Extraction}\label{FR}
We used EEG and eye-tracking to extract feature sets from navigational and informational components of each trial, enabling us to tailor our approach for optimal results.
\vspace{-0.2cm}
\paragraph{EEG based features}
% we use two different sets of features in our work: a set of features based on PyEEG~\cite{PyEEG}, as well as Common Spatial Pattern (CSP) features~\cite{DBLP:journals/nn/DaSallaKSK09}.
We used two different feature extraction methods for EEG signals. Firstly, we use PyEEG, an open-source Python module for EEG feature extraction~\cite{PyEEG} in the frequency and time domain. 
Additionally, we included statistical features, and in total, we extract $15$ features per channel, resulting in $960$ features ($64$ channels * $15$ features). Table~\ref{PyEEG-features} shows the extracted features for each EEG channel. 
As the number of PyEEG features is much larger than our extracted gaze features, we applied principal component analysis (PCA) for dimensionality reduction, similar to~\cite{slanzi2017combining}, we select the principal components where the explained variance is $90\%$. 
% resulting in $8$ to $25$ principal components.
Secondly, we used Common Spatial Pattern (CSP) to extract features from EEG data in a maximally discriminative manner~\cite{DBLP:journals/nn/DaSallaKSK09,DBLP:conf/premi/ChatterjeeBKTKJ13}. We used default parameters from the MNE toolbox~\cite{GramfortEtAl2013a}, resulting in $4$ spatial features.
 
%We also tried a threshold value of $t = 0.3$ where we get five features. However, the performance of $8$ features was better than $5$.}.

\vspace{-0.2cm}
\paragraph{Eye-tracking features}
%we compute eye-tracking features on the pre-processed data after applying fixation detection. 
Table~\ref{eye-features} shows the list of features  
adapted from the existing state-of-the-art~\cite{bhattacharya2020relevance,barz2021implicit}. Features are based on fixation events, saccadic eye movements, and the scanned area. Some features are time-normalized by the total time covered by the provided gaze data. We compute total$\_$time as the difference between the recording's last and first timestamp. 
After extracting the features, we follow a similar strategy as in~\cite{barz2021implicit} to reduce feature multi-collinearity by performing hierarchical clustering on the feature's Spearman rank-order correlations. We set the distance threshold to $0.2$ and use one feature per cluster
% \footnote{In practice, we set the distance threshold to $t = 0.2$.},
\sloppy
i.e., \texttt{fixn\_dur\_avg}, \texttt{fixn\_dur\_sd}, \texttt{scan\_hv\_ratio}, \texttt{fixn\_dur\_sum}, 
\texttt{scan\_speed\_h}, 
\texttt{fixns\_per\_box\_area}, \texttt{avg\_sacc\_length}, \texttt{scan\_speed\_v}.

% \vspace{-0.2cm}

\subsection{Classification and Fusion Techniques}
In our study, we use three popular classification algorithms, including Support Vector Machine (SVM), Random Forest (RF), and Na\"ive Bayes (NB), which are widely used in EEG- and eye-tracking studies~\cite{jang2014human, kang2015human,park2014human,zhao2020research}. %We applied these algorithms to both uni-modal and multi-modal prediction tasks. 
Further, we explored three distinct fusion mechanisms: early, late, and hybrid strategies. 
%Our preliminary experiments revealed that SVM with RBF kernel consistently performed better than other classifiers when employed at the final prediction stage. As a result we used SVM in all classification experiments. 
%Figure~\ref{fusion} visualizes all three techniques.
For every fusion approach, we first performed a min-max scaling on the input features to harmonize the scale of features within- and across modalities.

\vspace{-0.2cm}
\paragraph{Early fusion} This is the simple concatenation of EEG- and eye feature representations and inputs them as a single vector into the classifier.
%of the EEG and eye-tracking features before feeding it to the classifier. 
It aims to exploit dependencies between features. The input format from both modalities must be temporally compatible so that it is possible to combine them. 
%The data synchronization is performed based on the available timestamps.
This fusion technique is common in previous studies on navigational vs. informational intent classification~\cite{park2014human}.
% We chose SVM as a classifier for early fusion, as it performed consistently better than the alternatives in preliminary experiments.

% \begin{figure}
% \begin{center}
% \includegraphics[trim={0 5cm 0 1cm},clip, width=\linewidth]{pic/fusion_diagram.pdf}
% \end{center}
% \caption{Different fusion techniques: Early fusion focuses on feature level fusion while late, and Hybrid performs fusion at decision level}
% \label{fusion}
% \end{figure}

\vspace{-0.2cm}
\paragraph{Late fusion} This strategy trains classifiers for individual modalities - i.e., an EEG classifier with EEG features as input, and an eye tracking classifier with eye tracking features as input. To fuse the outputs of the modality-specific classifiers, we trained an additional classifier producing the final decision.

%This strategy aims first to train uni-modal classifiers independently, and their outputs are subsequently joined and fed to a second classifier that outputs the final decision. We choose the best classifier for each modality as our uni-modal classifiers. 
% \philipp{didn't we say this was always SVM? if we talk about choosing between different classifiers here, readers might be confused. Same for hybrid fusion paragraph.} 

%We chose SVM for the subsequent final classifier, which consistently yielded the best results in preliminary experiments.

\vspace{-0.2cm}
\paragraph{Hybrid fusion} The fusion happens at the shared representation layer from late and early fusion. It was successfully used in emotion recognition \cite{cimtay2020cross,nemati2019hybrid} and multimedia event detection \cite{lan2014multimedia}, to name a few. In addition to the two uni-modal classifier outputs in late fusion it also utilizes the output of the early fusion classifier described above. It then trains an additional classifier that predicts the final decision based on these three classifier outputs (uni-modal EEG, uni-modal eye, multimodal early fusion).

% We select the same models for the first and second stages of learning. However, instead of using two modalities, we only use predictions from the best-performing eye-tracking, EEG, and early fusion classifiers as input to our meta-model. 

%In preliminary experiments, we found SVM superior to RF or NB when applied at the final stage, so we always chose SVM for the meta-model.

%Figure~\ref{fusion} shows the hybrid fusion approach where fusion happens at the shared representation layer from late and early fusion.

For all fusion approaches and fusion stages, we evaluated different classifiers (SVM, NB, RF) in our experiments. SVM always performed best in these experiments, so we chose it as our classifier for all modalities and fusion stages.

% \vspace{-0.2cm}
\section{Evaluation}\label{res}
We present classification results for cross-user and user-specific prediction and evaluate different feature sets and multi-modal fusion methods.
In the cross-user scenario, we perform leave-one-user-out cross-validation. In the within-user scenario, we report mean accuracy averaged over $10$ random train-test splits to estimate the generalization performance for each user. For each train-test split, we perform feature selection (see Section~\ref{FR}) and parameter tuning via grid search cross-validation only on the train data and use the test set only for prediction. In all evaluation scenarios, the random baseline is at $50\%$ accuracy. % and we perform relevant statistical analysis to report the significance of the best-performing EEG feature method.

\begin{figure*}
\centering
\begin{subfigure}{.5\textwidth}
  \centering
  %{left bottom right top}
  \includegraphics[trim={0.5cm 0.55cm 1.5cm 1.2cm},clip,width=\linewidth]
  %{pic/bar_plot_svm_PyEEG_CSP_lopo.pdf}
  {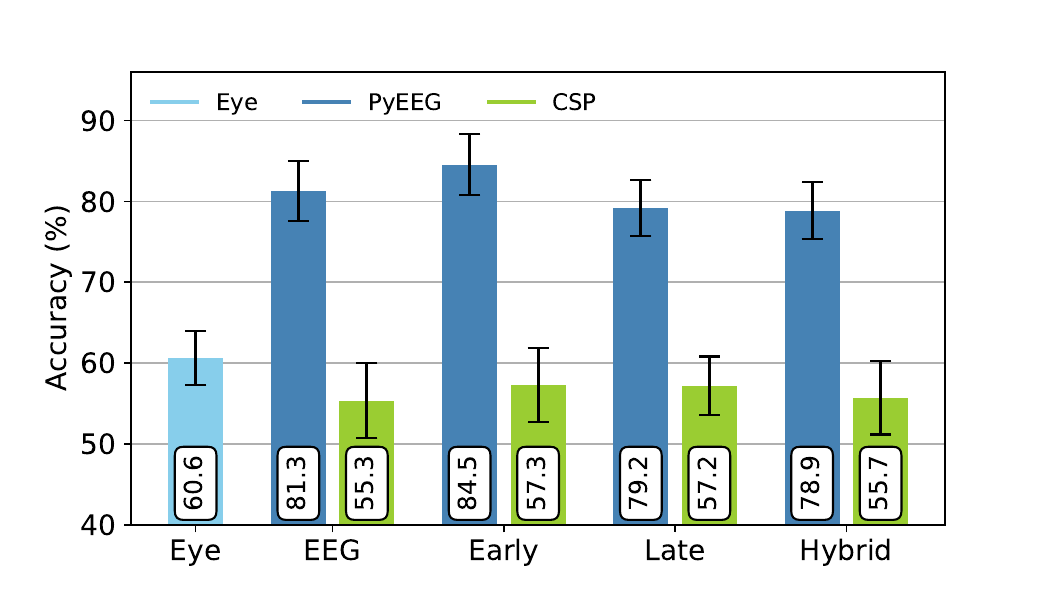}
  \caption{}
  \label{fig:main_results_a}
\end{subfigure}%
\begin{subfigure}{.5\textwidth}
  \centering
  \includegraphics[trim={0.5cm 0.60cm 1.5cm 1cm},clip,width=\linewidth]
  %{pic/bar_plot_svm_PyEEG_CSP_single.pdf}
  {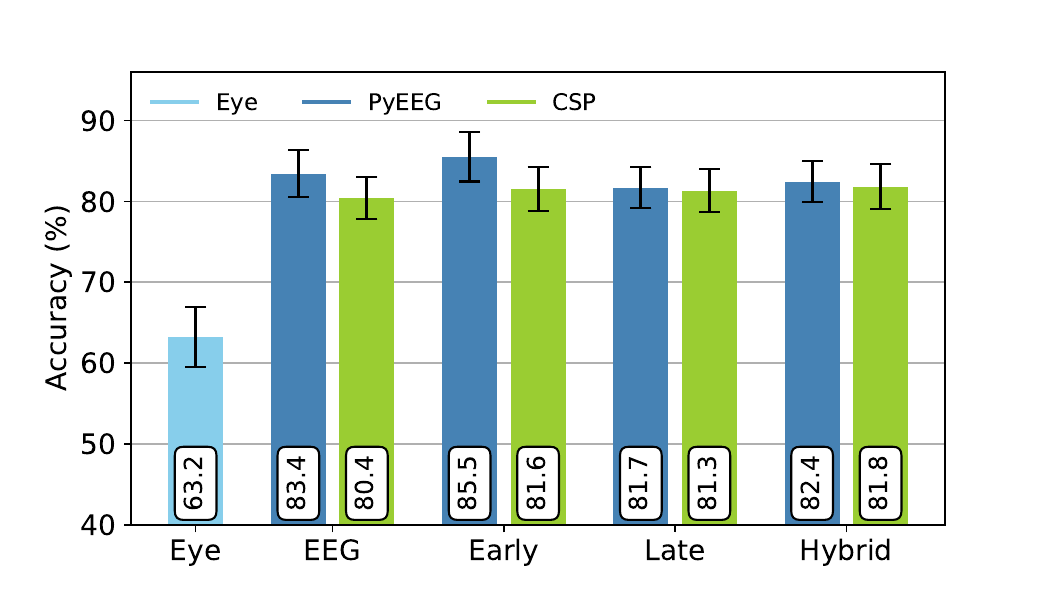}
  \caption{}
  \label{fig:main_results_b}
\end{subfigure}
\caption{(a) Comparison of modalities and fusion strategies for cross-user conditions using the best overall classifier (SVM)  (b) Comparison of modalities and fusion strategies for within-user conditions using the best overall classifier (SVM). Error bars indicate $95\%$ confidence interval.}%The bottom represents the mean accuracy. \philipp{what does this last sentence mean?}}
\label{fig:test}
\end{figure*}

% \vspace{-0.2cm}
\subsection{Cross-user Prediction}
We present cross-user prediction results for the leave-one-user-out evaluation scenario in \autoref{fig:main_results_a}.
The best approach relies on PyEEG and eye-tracking features joined by early fusion, reaching $84.5\%$ accuracy.
This is a significant increase over eye-tracking features with $60.6\%$ accuracy and a moderate increase over PyEEG-only features with $81.3\%$ accuracy, documenting the effectiveness of a multi-modal approach.
Interestingly, the CSP features performs far worse than the PyEEG dataset.
Unlike PyEEG, CSP performs below eye-tracking features, even in each multi-modal scenario.
There are no substantial differences between late and hybrid fusion strategies. Only early fusion showed an increase over the mono-modal EEG baseline. 
% For cross-user, all classifiers perform similarly well except low performance for Na\"ive Bayes, reaching $57.0\%$ and $65.0\%$ accuracy, while Random forest achieves $59.7\%$ and $77.4\%$ accuracy for eye and EEG, respectively. \philipp{don't describe the classifier comparisons here, rather describe what you show in the plots}
\autoref{fig:main_results_a} showed the error bars with $95\%$ confidence interval. We performed a paired t-test, and the calculated $p$-value of ~$6.1e-07$~ is less than the significance level $\alpha$ = $0.05$, showing that the choice of features (PyEEG vs. CSP) is significant. To compare the difference between the uni-modal Eye based classifier and multi-modal classifiers (Eye and PyEEG), we performed three pairwise t-test with the adjusted $\alpha$-value of $0.0167$ after Bonferroni correction. For eye, we obtained $p$-values of $4.8e-08$, $4.0e-07$, and $7.7e-08$ for early, late, and hybrid fusion, respectively, indicating significant improvements. Comparing uni-modal PyEEG features with multi-modal approaches (Eye and PyEEG), we obtained $p$-values of $0.02$, $0.26$, and $0.22$ for early, late, and hybrid fusion, respectively, showing no significant effect. We computed three paired t-tests for comparing different fusion methods with the adjusted $\alpha$ value of $0.167$ after Bonferroni correction. The results showed the difference between early and late fusion ($p$-value~=~$1.0e-03$), early and hybrid fusion ($p$-value = $6.6e-04$) is significant whereas late and hybrid fusion ($p$-value~=~$6.6e-1$) showed no significant effect.

\subsection{Within-user Prediction}
In \autoref{fig:main_results_b}, we present results for within-user prediction.
In line with the cross-user prediction scenario, the best method for within-user prediction was an early fusion of eye-tracking and PyEEG features, reaching $85.5\%$ accuracy.
This is a substantial increase over eye-tracking features with $63.2\%$ accuracy and a slight increase over PyEEG-only features with $83.4\%$ accuracy.
While the best method relies on joining eye-tracking- with PyEEG features, CSP features also perform well with an accuracy of $81.6\%$ for early fusion with eye-tracking features.
While there is an advantage for early fusion over late or hybrid fusion, the differences in fusion approaches are only moderate. 
% For within-user prediction, other classifiers follow the same pattern as cross-user. Random forest performs similarly to SVM, reaching $63.7\%$ and $77.7\%$ accuracy, while Na\"ive Bayes performs worse, achieving $59.5\%$ and $70.9\%$ accuracy for eye and EEG, respectively.    \philipp{these numbers on comparisons between classifiers are not in the plots, why do you describe them here?}
\autoref{fig:main_results_b} showed the error bars with $95\%$ confidence interval. We performed a paired t-test, and the calculated $p$-value of $2.0e-03$ is less than the significance level $\alpha$ = $0.05$. We have sufficient evidence that the choice of features (PyEEG vs. CSP) is significant.
To compare the difference between uni-modality (Eye, EEG-PyEEG) and multi-modality fusions, we performed a pairwise t-test with the adjusted $\alpha$-value of $0.0167$ after Bonferroni correction. For eye, we obtained $p$-values of $1.8e-29$, $1.8e-24$, and $2.0e-27$ for early, late, and hybrid fusion, respectively, showing that the improvements are highly significant. For EEG-PyEEG, we obtained $p$-values of $5.0e-04$, $4.0e-03$, and $0.15$ for early, late, and hybrid fusion, respectively, showing that the improvement for early and late fusion is highly significant. We computed three paired t-tests for comparing different fusion methods with the adjusted $\alpha$ value of $0.167$ after Bonferroni correction. The results showed the difference between early and late fusion ($p$-value = $1.1e-06$), early and hybrid fusion ($p$-value = $6.3e-05$) is significant whereas late and hybrid fusion ($p$-value = $2.9e-1$) showed no significant effect.

\subsection{Near Real-time Intent Prediction}
\begin{figure*}
\centering
\begin{subfigure}{.5\textwidth}
  \centering
  \includegraphics[trim={0.5cm 0.65cm 1.5cm 1cm},clip,width=\linewidth]
  %{pic/bar_plot_windows_lopo_l.pdf}
  {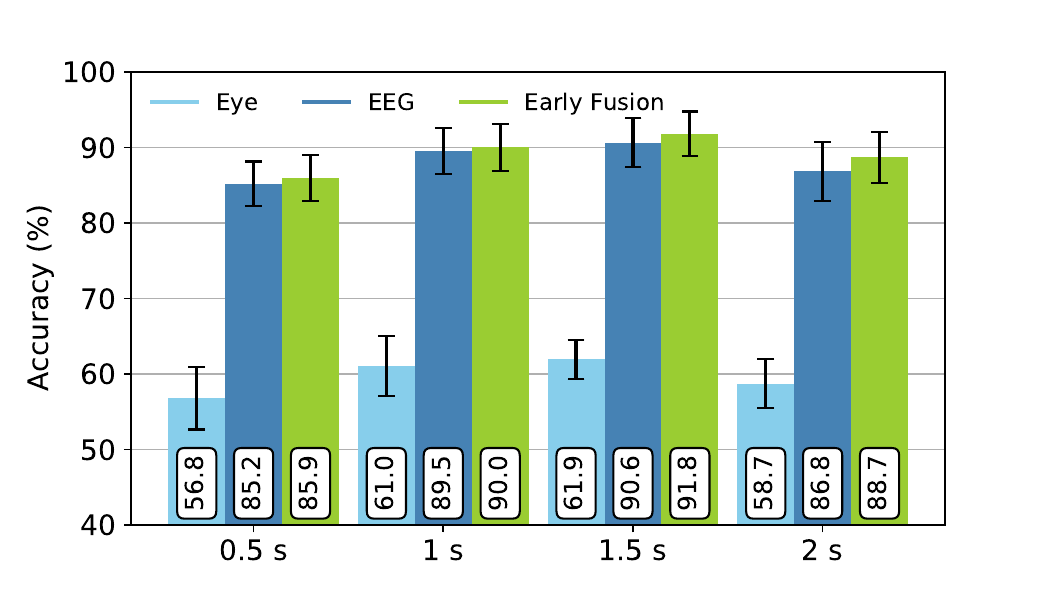}
  \caption{}
  \label{fig:cross_participant_window}
\end{subfigure}%
\begin{subfigure}{.5\textwidth}
  \centering
  \includegraphics[trim={0.5cm 0.65cm 1.5cm 1cm},clip,width=\linewidth]
  %{pic/bar_plot_windows_single_l.pdf}
  {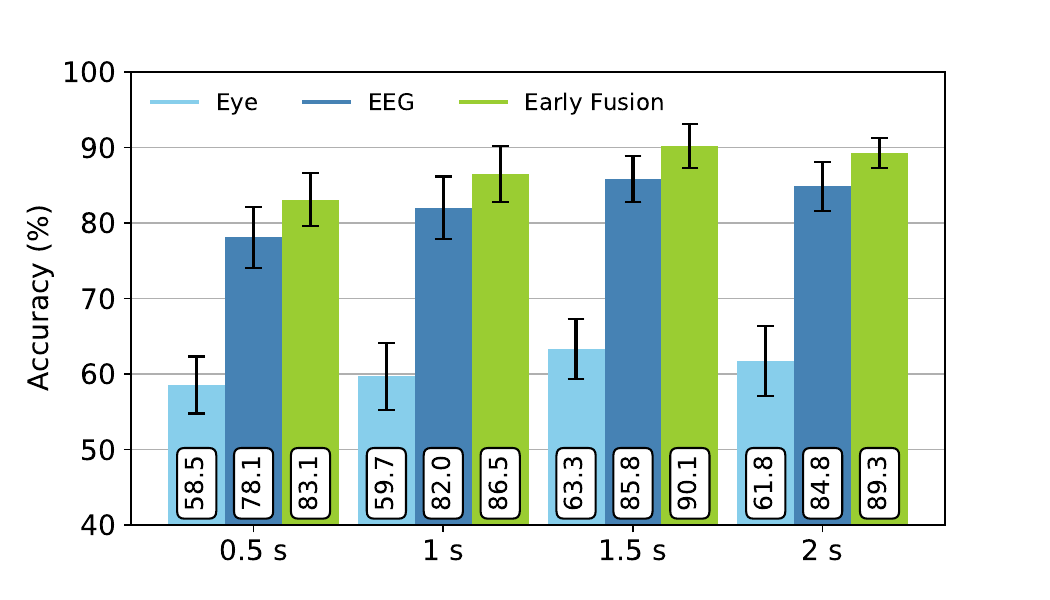}
  \caption{}
  \label{fig:within_participant_window}
\end{subfigure}
\caption{(a) Comparison with shorter time windows in cross-user condition with the best classifier (SVM), feature extraction (PyEEG), and fusion method (Early). (b) Comparison with shorter time windows in within-user condition with the best classifier (SVM), feature extraction (PyEEG), and fusion method (Early). Error bars indicate $95\%$ confidence interval.} % The bottom represents the mean accuracy.}
\label{fig:per_participant_windows}
\end{figure*}

We evaluate the influence of smaller search durations on user performance to understand better whether our approach can be used in near real-time intent prediction with the best configuration of EEG features and fusion methods. We select four different window sizes, %for this analysis
from $0.5$s to $2$s, as post $2$s duration, most of the users can locate the target, see~\autoref{fig:search_time} (left). Monitoring the search performance and enabling proactive support to minimize search delays in real-time would be beneficial within these window sizes. To make a fair comparison across different window sizes, we take the same samples and therefore exclude samples where the search duration is less than $2$s. 
Figure~\ref{fig:cross_participant_window} showed the results for cross-user prediction, where $1.5$s window achieves the best mean accuracy of $91.8\%$. Moreover, within-user follows a similar trend, as shown in \autoref{fig:within_participant_window}, achieving the best mean accuracy of $90.1\%$. Interestingly, early fusion improves over mono-modal inputs in both prediction scenarios across all window sizes, with a much higher increase in within-user conditions. To compare the difference between uni-modality (Eye, EEG-PyEEG) and multi-modality fusions for the best-performing window size of $1.5$s and for cross-user prediction, we performed a pairwise t-test with the $\alpha$-value of $0.05$. We obtained $p$-values of $1.0e-09$ and $0.18$ for eye and EEG, respectively. Therefore, the improvement of early fusion over eye-tracking is highly significant.
For similar analysis in the case of within-user prediction, we obtained $p$-values of $1.9e-31$ and $3.2e-10$ for eye and EEG, respectively. Therefore, the improvement of early fusion over eye-tracking and EEG is highly significant.
  
% Enough information to make accurate prediction, deploy model  in real-world, within a latency of 1.5s approx. 90% accuracy we can differentiate if user is looking around or searching. Even in 0.5s, model can predict with almost 85% intent, which gets further strengthened with the first 1.5s making it promising and feasible to deploy in real-world scenarios
% \Mansi{Result}
% \philipp{describe results + be cautious with using ``significant''. here you could just omit it and talk about a larger increase.}
%  write mean accuracy
% \begin{figure}
% \begin{center}
% \includegraphics[{0.5cm 0cm 5cm 0cm},clip,width=1.1\linewidth]{pic/bar_plot_windows_lopo_l.pdf}
% \end{center}
% \vspace{-0.4cm}
% \caption{Comparison with shorter time windows in cross-user condition with the best classifier (SVM), feature extraction (PyEEG), and fusion method (Early). Error bars indicate $95\%$ confidence interval. The bottom represents the mean accuracy.}
% \label{fig:cross_participant_window}
% \end{figure}

% \begin{figure}
% \begin{center}
% \includegraphics[{0.5cm 0cm 5cm 0cm},clip,width=1.1\linewidth]{pic/bar_plot_windows_single_l.pdf}
% \end{center}
% \vspace{-0.4cm}
% \caption{Comparison with shorter time windows in user-specific condition with the best classifier (SVM), feature extraction (PyEEG), and fusion method (Early). Error bars indicate $95\%$ confidence interval. The bottom represents the mean accuracy.}
% \label{fig:per_participant_windows}
% \end{figure}

% \vspace{-0.1cm}
\section{Discussion}
In the following section, we discuss the obtained results and focus on our method's performance and potential applications.

% \vspace{-0.2cm}
\subsection{Achieved Performance}
We presented the first method to the best of our knowledge for cross-user prediction of navigational vs. informational intent, reaching an accuracy of $84.5\%$.
While this is comparable to the accuracy we reached for within-user prediction $85.5\%$, it is a highly promising result, allowing for much larger flexibility in application scenarios.
In particular, when deploying an intent prediction system in an industrial context, collecting user-specific training data might not be feasible, highlighting the importance of effective cross-user prediction. %We observed that the choice of feature set is key to making cross-user prediction work. 
Using the CSP features extracted from EEG, we could not reach accuracies beyond $60\%$. With PyEEG features and feature selection, however, we reached a much higher accuracy of $84.5\%$ when combined with eye-tracking features.
There was no such substantial difference between EEG feature sets for within-user prediction.
Thus, a key takeaway from our results is that cross-user prediction requires careful feature selection.
%, and features that work in a within-user scenario might not generalize across users.

% \vspace{-0.2cm}
\subsection{Applications}

The ability to infer users' search intents by combining eye-tracking with the EEG data helps researchers study perceptual, attentional, or cognitive processes in more realistic situations as it aims to understand users' search intents without the need for them to communicate these intents verbally. By presenting the first method for cross-user prediction of navigational vs. informational intent, we pave the way for exciting new applications in other domains, such as VR-based gaming, where it can be used to provide a more immersive and interactive experience. For example, if the user searches for an object in a game, the system can adjust the game environment to make it more challenging based on the user's intent. The prime motivation of our work is applications in industrial work scenarios where workers are often faced with the task of finding a tool in a cluttered scene. Once informational intent is recognized for a user, a support system can help to find the desired tool. To know which tool the user is looking for, at least two approaches are conceivable. First, the support system could maintain a model of the user's current task and infer the tool the user most likely needs next to accomplish their task.
Second, the prediction of informational intent could be followed by search target prediction~\cite{sattar2015prediction,barz2020visual}, informing the system about the tool the user is looking for.
Apart from directly supporting humans in visual search, informational vs. navigational intent prediction could also be used to quantify how much time users spend in visual search.
Repeated long search times may indicate that the work environment needs to be simplified or tidy for efficient work.
% Based on these observations, organizations could take a targeted approach to re-structure or simplify specific work environments to minimize the time spent in visual search.

% \vspace{-0.4cm}
\subsection{Limitations}
While our novel dataset and cross-user prediction approach represent a significant step towards recognizing navigational and informational intent in the real world, some limitations remain.
%and opportunities for future work remain.
While using virtually created work scenes allowed us to include a large variety of visual environments in our data collection, a laboratory experiment always implies a domain gap to the real world. Especially the analysis of EEG signals is challenging in real-world environments due to motion artifacts and noise~\cite{chi2011dry}. However, researchers are developing wearable dry EEG electrodes devices to improve the overall reliability and applicability of EEG signal analysis in real-world scenarios and advanced signal processing techniques like continuous contact impedance monitoring~\cite{bertrand2013motion} and Gaussian Elimination Canonical Correlation Analysis~\cite{roy2017gaussian}.
Although we utilized classical classification and multi-modal fusion techniques to distinguish between navigational and informational intent recognition, which were widely used in the current state-of-the-art, given the size of the dataset, it would be interesting to observe the performance of deep learning architecture on this dataset.

\section{Conclusion and Future Work}\label{conclusion}

% In this work, we presented the first publicly available EEG- and eye-tracking dataset for informational versus navigational intent recognition.
% Our dataset improves over previous recording procedures by using a concrete application scenario with fully user-defined search times.
% By making the data publicly available, we hope to increase the attention of the research community on this important intent recognition task.
% In addition, we presented the first method for navigational versus informational intent recognition from EEG and eye-tracking that applies to users unseen at test time - an essential requirement in many practical settings.
% We uncover that careful selection of features is crucial in this challenging prediction scenario.
% Taken together, our work significantly advances informational versus navigational intent prediction in real-world scenarios.
We proposed the first publicly available EEG- and eye-tracking dataset for informational vs. navigational intent recognition and a multi-modal approach for classifying intents for within-user and leave-one-user-out scenarios.
Our dataset improves over previous recording procedures using a concrete application scenario with fully user-defined search times.
% By making the data publicly available, we aim to increase the attention of the research community on this vital intent recognition task. 
We thoroughly analyzed the dataset by evaluating it for within-user and cross-user scenarios and comparing different shorter time windows to demonstrate the potential for near real-time intent recognition. The presented statistical analysis highlights the importance of selecting appropriate features and fusion methods. %Overall, this study represents a significant advancement in the field of navigational vs. informational intent prediction in real-world scenarios. 
Future work should consider different scenarios to predict informational and navigational intent, including hospitals, retail, and even people's private spaces. Even after recognizing navigational vs. informational intent, the identity of the object the user searches for is not apparent. For such scenarios, future work should investigate how to integrate navigational and informational intent classification with search target prediction to help the user and reduce search times most effectively.
%in several application scenarios 

\begin{acks}
This work is funded by the \grantsponsor{01IW20003}{German Ministry for Education and Research (BMBF)}{https://www.bmbf.de/}, grant number \grantnum{}{01IW20003}. Philipp M\"uller is funded by the \grantsponsor{01IS20075}{German Ministry for Education and Research (BMBF)}{https://www.bmbf.de/}, grant number \grantnum{}{01IS20075}.
\end{acks}

%\Mansi{check if order of figures/table is fine}
\bibliographystyle{ACM-Reference-Format}
\bibliography{sample-base}


\begin{thebibliography}{41}

%%% the Web address.

\ifx \showCODEN    \undefined \def \showCODEN     #1{\unskip}     \fi
\ifx \showDOI      \undefined \def \showDOI       #1{#1}\fi
\ifx \showISBNx    \undefined \def \showISBNx     #1{\unskip}     \fi
\ifx \showISBNxiii \undefined \def \showISBNxiii  #1{\unskip}     \fi
\ifx \showISSN     \undefined \def \showISSN      #1{\unskip}     \fi
\ifx \showLCCN     \undefined \def \showLCCN      #1{\unskip}     \fi
\ifx \shownote     \undefined \def \shownote      #1{#1}          \fi
\ifx \showarticletitle \undefined \def \showarticletitle #1{#1}   \fi
\ifx \showURL      \undefined \def \showURL       {\relax}        \fi
% invisible to TeX
\providecommand\bibfield[2]{#2}
\providecommand\bibinfo[2]{#2}
\providecommand\natexlab[1]{#1}
\providecommand\showeprint[2][]{arXiv:#2}

\bibitem[Acharya et~al\mbox{.}(2016)]%
        {electrodes}
\bibfield{author}{\bibinfo{person}{Jayant Acharya}, \bibinfo{person}{Abeer
  Hani}, \bibinfo{person}{Janna Cheek}, \bibinfo{person}{Parthasarathy
  Thirumala}, {and} \bibinfo{person}{Tammy Tsuchida}.}
  \bibinfo{year}{2016}\natexlab{}.
\newblock \showarticletitle{American Clinical Neurophysiology Society Guideline
  2: Guidelines for Standard Electrode Position Nomenclature}.
\newblock \bibinfo{journal}{\emph{The Neurodiagnostic Journal}}
  \bibinfo{volume}{56} (\bibinfo{date}{10} \bibinfo{year}{2016}),
  \bibinfo{pages}{245--252}.
\newblock
\urldef\tempurl%
\url{https://doi.org/10.1080/21646821.2016.1245558}
\showDOI{\tempurl}


\bibitem[Bao et~al\mbox{.}(2011)]%
        {PyEEG}
\bibfield{author}{\bibinfo{person}{Forrest~Sheng Bao}, \bibinfo{person}{Xin
  Liu}, {and} \bibinfo{person}{Christina Zhang}.}
  \bibinfo{year}{2011}\natexlab{}.
\newblock \showarticletitle{PyEEG: An Open Source Python Module for {EEG/MEG}
  Feature Extraction}.
\newblock \bibinfo{journal}{\emph{Comp. Int. and Neurosc.}}
  (\bibinfo{year}{2011}), \bibinfo{pages}{406391:1--406391:7}.
\newblock
\urldef\tempurl%
\url{https://doi.org/10.1155/2011/406391}
\showDOI{\tempurl}


\bibitem[Barz et~al\mbox{.}(2022)]%
        {barz2021implicit}
\bibfield{author}{\bibinfo{person}{Michael Barz},
  \bibinfo{person}{Omair~Shahzad Bhatti}, {and} \bibinfo{person}{Daniel
  Sonntag}.} \bibinfo{year}{2022}\natexlab{}.
\newblock \showarticletitle{Implicit Estimation of Paragraph Relevance From Eye
  Movements}.
\newblock \bibinfo{journal}{\emph{Frontiers in Computer Science}}
  \bibinfo{volume}{3} (\bibinfo{year}{2022}).
\newblock
\showISSN{2624-9898}
\urldef\tempurl%
\url{https://doi.org/10.3389/fcomp.2021.808507}
\showDOI{\tempurl}


\bibitem[Barz et~al\mbox{.}(2020)]%
        {barz2020visual}
\bibfield{author}{\bibinfo{person}{Michael Barz}, \bibinfo{person}{Sven
  Stauden}, {and} \bibinfo{person}{Daniel Sonntag}.}
  \bibinfo{year}{2020}\natexlab{}.
\newblock \showarticletitle{Visual Search Target Inference in Natural
  Interaction Settings with Machine Learning}. In \bibinfo{booktitle}{\emph{ACM
  Symposium on Eye Tracking Research and Applications}}
  \emph{(\bibinfo{series}{ETRA '20 Full Papers})}. Article
  \bibinfo{articleno}{1}, \bibinfo{numpages}{8}~pages.
\newblock
\urldef\tempurl%
\url{https://doi.org/10.1145/3379155.3391314}
\showDOI{\tempurl}


\bibitem[Bertrand et~al\mbox{.}(2013)]%
        {bertrand2013motion}
\bibfield{author}{\bibinfo{person}{Alexander Bertrand}, \bibinfo{person}{Vojkan
  Mihajlović}, \bibinfo{person}{Bernard Grundlehner}, \bibinfo{person}{Chris
  Van~Hoof}, {and} \bibinfo{person}{Marc Moonen}.}
  \bibinfo{year}{2013}\natexlab{}.
\newblock \showarticletitle{Motion artifact reduction in EEG recordings using
  multi-channel contact impedance measurements}. In
  \bibinfo{booktitle}{\emph{2013 IEEE Biomedical Circuits and Systems
  Conference (BioCAS)}}. \bibinfo{pages}{258--261}.
\newblock
\urldef\tempurl%
\url{https://doi.org/10.1109/BioCAS.2013.6679688}
\showDOI{\tempurl}


\bibitem[Bhattacharya et~al\mbox{.}(2020)]%
        {bhattacharya2020relevance}
\bibfield{author}{\bibinfo{person}{Nilavra Bhattacharya},
  \bibinfo{person}{Somnath Rakshit}, \bibinfo{person}{Jacek Gwizdka}, {and}
  \bibinfo{person}{Paul Kogut}.} \bibinfo{year}{2020}\natexlab{}.
\newblock \showarticletitle{Relevance Prediction from Eye-Movements Using
  Semi-Interpretable Convolutional Neural Networks}. In
  \bibinfo{booktitle}{\emph{Proceedings of the 2020 Conference on Human
  Information Interaction and Retrieval}}. \bibinfo{pages}{223–233}.
\newblock
\urldef\tempurl%
\url{https://doi.org/10.1145/3343413.3377960}
\showDOI{\tempurl}


\bibitem[Chatterjee et~al\mbox{.}(2013)]%
        {DBLP:conf/premi/ChatterjeeBKTKJ13}
\bibfield{author}{\bibinfo{person}{Soumyadip Chatterjee},
  \bibinfo{person}{Saugat Bhattacharyya}, \bibinfo{person}{Amit Konar},
  \bibinfo{person}{D.~N. Tibarewala}, \bibinfo{person}{Anwesha Khasnobish},
  {and} \bibinfo{person}{Ramadoss Janarthanan}.}
  \bibinfo{year}{2013}\natexlab{}.
\newblock \showarticletitle{Performance Analysis of Multiclass Common Spatial
  Patterns in Brain-Computer Interface}. In \bibinfo{booktitle}{\emph{Pattern
  Recognition and Machine Intelligence - 5th International Conference, PReMI
  2013}}. \bibinfo{pages}{115--120}.
\newblock
\urldef\tempurl%
\url{https://doi.org/10.1007/978-3-642-45062-4\_15}
\showDOI{\tempurl}


\bibitem[Cheng et~al\mbox{.}(2020)]%
        {Shiweinew}
\bibfield{author}{\bibinfo{person}{Shiwei Cheng}, \bibinfo{person}{Jialing
  Wang}, \bibinfo{person}{Lekai Zhang}, {and} \bibinfo{person}{Qianjing Wei}.}
  \bibinfo{year}{2020}\natexlab{}.
\newblock \showarticletitle{Motion Imagery-BCI Based on EEG and Eye Movement
  Data Fusion}.
\newblock \bibinfo{journal}{\emph{IEEE Transactions on Neural Systems and
  Rehabilitation Engineering}}  \bibinfo{volume}{PP} (\bibinfo{date}{12}
  \bibinfo{year}{2020}), \bibinfo{pages}{1--1}.
\newblock
\urldef\tempurl%
\url{https://doi.org/10.1109/TNSRE.2020.3048422}
\showDOI{\tempurl}


\bibitem[Chi et~al\mbox{.}(2012)]%
        {chi2011dry}
\bibfield{author}{\bibinfo{person}{Yu~Mike Chi}, \bibinfo{person}{Yu-Te Wang},
  \bibinfo{person}{Yijun Wang}, \bibinfo{person}{Christoph Maier},
  \bibinfo{person}{Tzyy-Ping Jung}, {and} \bibinfo{person}{Gert Cauwenberghs}.}
  \bibinfo{year}{2012}\natexlab{}.
\newblock \showarticletitle{Dry and Noncontact EEG Sensors for Mobile
  Brain–Computer Interfaces}.
\newblock \bibinfo{journal}{\emph{IEEE Transactions on Neural Systems and
  Rehabilitation Engineering}} \bibinfo{volume}{20}, \bibinfo{number}{2}
  (\bibinfo{year}{2012}), \bibinfo{pages}{228--235}.
\newblock
\urldef\tempurl%
\url{https://doi.org/10.1109/TNSRE.2011.2174652}
\showDOI{\tempurl}


\bibitem[Cimtay et~al\mbox{.}(2020)]%
        {cimtay2020cross}
\bibfield{author}{\bibinfo{person}{Yucel Cimtay}, \bibinfo{person}{Erhan
  Ekmekcioglu}, {and} \bibinfo{person}{Seyma Caglar-Ozhan}.}
  \bibinfo{year}{2020}\natexlab{}.
\newblock \showarticletitle{Cross-Subject Multimodal Emotion Recognition Based
  on Hybrid Fusion}.
\newblock \bibinfo{journal}{\emph{IEEE Access}}  \bibinfo{volume}{8}
  (\bibinfo{year}{2020}), \bibinfo{pages}{168865--168878}.
\newblock
\urldef\tempurl%
\url{https://doi.org/10.1109/ACCESS.2020.3023871}
\showDOI{\tempurl}


\bibitem[DaSalla et~al\mbox{.}(2009)]%
        {DBLP:journals/nn/DaSallaKSK09}
\bibfield{author}{\bibinfo{person}{Charles~S. DaSalla},
  \bibinfo{person}{Hiroyuki Kambara}, \bibinfo{person}{Makoto Sato}, {and}
  \bibinfo{person}{Yasuharu Koike}.} \bibinfo{year}{2009}\natexlab{}.
\newblock \showarticletitle{Single-trial classification of vowel speech imagery
  using common spatial patterns}.
\newblock \bibinfo{journal}{\emph{Neural Networks}} \bibinfo{volume}{22},
  \bibinfo{number}{9} (\bibinfo{year}{2009}), \bibinfo{pages}{1334--1339}.
\newblock
\urldef\tempurl%
\url{https://doi.org/10.1016/j.neunet.2009.05.008}
\showDOI{\tempurl}


\bibitem[Delorme and Makeig(2004)]%
        {delorme2004eeglab}
\bibfield{author}{\bibinfo{person}{Arnaud Delorme} {and} \bibinfo{person}{Scott
  Makeig}.} \bibinfo{year}{2004}\natexlab{}.
\newblock \showarticletitle{EEGLAB: an open source toolbox for analysis of
  single-trial EEG dynamics including independent component analysis}.
\newblock \bibinfo{journal}{\emph{Journal of Neuroscience Methods}}
  \bibinfo{volume}{134}, \bibinfo{number}{1} (\bibinfo{year}{2004}),
  \bibinfo{pages}{9--21}.
\newblock
\urldef\tempurl%
\url{https://doi.org/10.1016/j.jneumeth.2003.10.009}
\showDOI{\tempurl}


\bibitem[Ferree et~al\mbox{.}(2001)]%
        {imp}
\bibfield{author}{\bibinfo{person}{Thomas~C Ferree}, \bibinfo{person}{Phan
  Luu}, \bibinfo{person}{Gerald~S Russell}, {and} \bibinfo{person}{Don~M
  Tucker}.} \bibinfo{year}{2001}\natexlab{}.
\newblock \showarticletitle{Scalp electrode impedance, infection risk, and EEG
  data quality}.
\newblock \bibinfo{journal}{\emph{Clinical Neurophysiology}}
  \bibinfo{volume}{112}, \bibinfo{number}{3} (\bibinfo{year}{2001}),
  \bibinfo{pages}{536--544}.
\newblock
\urldef\tempurl%
\url{https://doi.org/10.1016/S1388-2457(00)00533-2}
\showDOI{\tempurl}


\bibitem[Gramfort et~al\mbox{.}(2013)]%
        {GramfortEtAl2013a}
\bibfield{author}{\bibinfo{person}{Alexandre Gramfort}, \bibinfo{person}{Martin
  Luessi}, \bibinfo{person}{Eric Larson}, \bibinfo{person}{Denis~A. Engemann},
  \bibinfo{person}{Daniel Strohmeier}, \bibinfo{person}{Christian Brodbeck},
  \bibinfo{person}{Roman Goj}, \bibinfo{person}{Mainak Jas},
  \bibinfo{person}{Teon Brooks}, \bibinfo{person}{Lauri Parkkonen}, {and}
  \bibinfo{person}{Matti~S. H{\"a}m{\"a}l{\"a}inen}.}
  \bibinfo{year}{2013}\natexlab{}.
\newblock \showarticletitle{{{MEG}} and {{EEG}} Data Analysis with
  {{MNE}}-{{Python}}}.
\newblock \bibinfo{journal}{\emph{Frontiers in Neuroscience}}
  \bibinfo{volume}{7}, \bibinfo{number}{267} (\bibinfo{year}{2013}),
  \bibinfo{pages}{1--13}.
\newblock
\urldef\tempurl%
\url{https://doi.org/10.3389/fnins.2013.00267}
\showDOI{\tempurl}


\bibitem[Haas(2014)]%
        {unity}
\bibfield{author}{\bibinfo{person}{John~K Haas}.}
  \bibinfo{year}{2014}\natexlab{}.
\newblock \showarticletitle{A History of the Unity Game Engine}.
\newblock
\urldef\tempurl%
\url{https://api.semanticscholar.org/CorpusID:86824974}
\showURL{%
\tempurl}


\bibitem[Huang et~al\mbox{.}(2015)]%
        {huang2015using}
\bibfield{author}{\bibinfo{person}{Chien-Ming Huang}, \bibinfo{person}{Sean
  Andrist}, \bibinfo{person}{Allison Sauppé}, {and} \bibinfo{person}{Bilge
  Mutlu}.} \bibinfo{year}{2015}\natexlab{}.
\newblock \showarticletitle{Using gaze patterns to predict task intent in
  collaboration}.
\newblock \bibinfo{journal}{\emph{Frontiers in Psychology}}
  \bibinfo{volume}{6} (\bibinfo{year}{2015}).
\newblock
\urldef\tempurl%
\url{https://doi.org/10.3389/fpsyg.2015.01049}
\showDOI{\tempurl}


\bibitem[Jang et~al\mbox{.}(2014a)]%
        {jang2014identification}
\bibfield{author}{\bibinfo{person}{Young-Min Jang}, \bibinfo{person}{Rammohan
  Mallipeddi}, {and} \bibinfo{person}{Minho Lee}.}
  \bibinfo{year}{2014}\natexlab{a}.
\newblock \showarticletitle{Identification of human implicit visual search
  intention based on eye movement and pupillary analysis}.
\newblock \bibinfo{journal}{\emph{User Modeling and User-Adapted Interaction}}
  \bibinfo{volume}{24}, \bibinfo{number}{4} (\bibinfo{year}{2014}),
  \bibinfo{pages}{315--344}.
\newblock
\urldef\tempurl%
\url{https://doi.org/doi/10.1007/s11257-013-9142-7}
\showDOI{\tempurl}


\bibitem[Jang et~al\mbox{.}(2014b)]%
        {jang2014human}
\bibfield{author}{\bibinfo{person}{Young-Min Jang}, \bibinfo{person}{Rammohan
  Mallipeddi}, \bibinfo{person}{Sangil Lee}, \bibinfo{person}{Ho-Wan Kwak},
  {and} \bibinfo{person}{Minho Lee}.} \bibinfo{year}{2014}\natexlab{b}.
\newblock \showarticletitle{Human intention recognition based on eyeball
  movement pattern and pupil size variation}.
\newblock \bibinfo{journal}{\emph{Neurocomputing}}  \bibinfo{volume}{128}
  (\bibinfo{year}{2014}), \bibinfo{pages}{421--432}.
\newblock
\showISSN{0925-2312}
\urldef\tempurl%
\url{https://doi.org/10.1016/j.neucom.2013.08.008}
\showDOI{\tempurl}


\bibitem[Kang et~al\mbox{.}(2015)]%
        {kang2015human}
\bibfield{author}{\bibinfo{person}{Jun-Su Kang}, \bibinfo{person}{Ukeob Park},
  \bibinfo{person}{V. Gonuguntla}, \bibinfo{person}{K.C. Veluvolu}, {and}
  \bibinfo{person}{Minho Lee}.} \bibinfo{year}{2015}\natexlab{}.
\newblock \showarticletitle{Human implicit intent recognition based on the
  phase synchrony of EEG signals}.
\newblock \bibinfo{journal}{\emph{Pattern Recognition Letters}}
  \bibinfo{volume}{66} (\bibinfo{year}{2015}), \bibinfo{pages}{144--152}.
\newblock
\urldef\tempurl%
\url{https://doi.org/10.1016/j.patrec.2015.06.013}
\showDOI{\tempurl}
\newblock
\shownote{Pattern Recognition in Human Computer Interaction}.


\bibitem[Klug and Gramann(2021)]%
        {klug2021identifying}
\bibfield{author}{\bibinfo{person}{Marius Klug} {and} \bibinfo{person}{Klaus
  Gramann}.} \bibinfo{year}{2021}\natexlab{}.
\newblock \showarticletitle{Identifying key factors for improving ICA-based
  decomposition of EEG data in mobile and stationary experiments}.
\newblock \bibinfo{journal}{\emph{European Journal of Neuroscience}}
  \bibinfo{volume}{54}, \bibinfo{number}{12} (\bibinfo{year}{2021}),
  \bibinfo{pages}{8406--8420}.
\newblock
\urldef\tempurl%
\url{https://doi.org/10.1111/ejn.14992}
\showDOI{\tempurl}


\bibitem[Lan et~al\mbox{.}(2014)]%
        {lan2014multimedia}
\bibfield{author}{\bibinfo{person}{Zhen-zhong Lan}, \bibinfo{person}{Lei Bao},
  \bibinfo{person}{Shoou-I Yu}, \bibinfo{person}{Wei Liu}, {and}
  \bibinfo{person}{Alexander~G Hauptmann}.} \bibinfo{year}{2014}\natexlab{}.
\newblock \showarticletitle{Multimedia classification and event detection using
  double fusion}.
\newblock \bibinfo{journal}{\emph{Multimedia tools and applications}}
  \bibinfo{volume}{71}, \bibinfo{number}{1} (\bibinfo{year}{2014}),
  \bibinfo{pages}{333--347}.
\newblock
\urldef\tempurl%
\url{https://doi.org/10.1007/s11042-013-1391-2}
\showDOI{\tempurl}


\bibitem[Levy(1987)]%
        {levy1987effect}
\bibfield{author}{\bibinfo{person}{Warren~J Levy}.}
  \bibinfo{year}{1987}\natexlab{}.
\newblock \showarticletitle{Effect of epoch length on power spectrum analysis
  of the EEG.}
\newblock \bibinfo{journal}{\emph{Anesthesiology}} \bibinfo{volume}{66},
  \bibinfo{number}{4} (\bibinfo{year}{1987}), \bibinfo{pages}{489--495}.
\newblock
\urldef\tempurl%
\url{https://doi.org/10.1097/00000542-198704000-00007}
\showDOI{\tempurl}


\bibitem[Liang et~al\mbox{.}(2019)]%
        {liang2019application}
\bibfield{author}{\bibinfo{person}{Yongqiang Liang}, \bibinfo{person}{Wei
  Wang}, \bibinfo{person}{Jue Qu}, {and} \bibinfo{person}{Jie Yang}.}
  \bibinfo{year}{2019}\natexlab{}.
\newblock \showarticletitle{Application of Eye Tracking in Intelligent User
  Interface}.
\newblock \bibinfo{journal}{\emph{Journal of Physics: Conference Series}}
  \bibinfo{volume}{1169}, \bibinfo{number}{1} (\bibinfo{date}{feb}
  \bibinfo{year}{2019}), \bibinfo{pages}{012040}.
\newblock
\urldef\tempurl%
\url{https://doi.org/10.1088/1742-6596/1169/1/012040}
\showDOI{\tempurl}


\bibitem[Ludwig et~al\mbox{.}(2009)]%
        {ludwig2009using}
\bibfield{author}{\bibinfo{person}{Kip Ludwig}, \bibinfo{person}{Rachel
  Miriani}, \bibinfo{person}{Nicholas Langhals}, \bibinfo{person}{Michael
  Joseph}, \bibinfo{person}{David Anderson}, {and} \bibinfo{person}{Daryl
  Kipke}.} \bibinfo{year}{2009}\natexlab{}.
\newblock \showarticletitle{Using a Common Average Reference to Improve
  Cortical Neuron Recordings From Microelectrode Arrays}.
\newblock \bibinfo{journal}{\emph{Journal of neurophysiology}}
  \bibinfo{volume}{101} (\bibinfo{date}{03} \bibinfo{year}{2009}),
  \bibinfo{pages}{1679--89}.
\newblock
\urldef\tempurl%
\url{https://doi.org/10.1152/jn.90989.2008}
\showDOI{\tempurl}


\bibitem[Majaranta and Bulling(2014)]%
        {majaranta2014eye}
\bibfield{author}{\bibinfo{person}{P{\"a}ivi Majaranta} {and}
  \bibinfo{person}{Andreas Bulling}.} \bibinfo{year}{2014}\natexlab{}.
\newblock \bibinfo{booktitle}{\emph{Eye Tracking and Eye-Based Human--Computer
  Interaction}}.
\newblock \bibinfo{pages}{39--65}.
\newblock
\urldef\tempurl%
\url{https://doi.org/10.1007/978-1-4471-6392-3_3}
\showDOI{\tempurl}


\bibitem[Müller et~al\mbox{.}(2020)]%
        {mueller20_etra}
\bibfield{author}{\bibinfo{person}{Philipp Müller}, \bibinfo{person}{Ekta
  Sood}, {and} \bibinfo{person}{Andreas Bulling}.}
  \bibinfo{year}{2020}\natexlab{}.
\newblock \showarticletitle{Anticipating Averted Gaze in Dyadic Interactions}.
  In \bibinfo{booktitle}{\emph{Proceedings of ACM International Symposium on
  Eye Tracking Research and Applications (ETRA)}}. \bibinfo{pages}{1--10}.
\newblock
\urldef\tempurl%
\url{https://doi.org/10.1145/3379155.3391332}
\showDOI{\tempurl}


\bibitem[Nemati et~al\mbox{.}(2019)]%
        {nemati2019hybrid}
\bibfield{author}{\bibinfo{person}{Shahla Nemati}, \bibinfo{person}{Reza
  Rohani}, \bibinfo{person}{Mohammad~Ehsan Basiri}, \bibinfo{person}{Moloud
  Abdar}, \bibinfo{person}{Neil~Y. Yen}, {and} \bibinfo{person}{Vladimir
  Makarenkov}.} \bibinfo{year}{2019}\natexlab{}.
\newblock \showarticletitle{A Hybrid Latent Space Data Fusion Method for
  Multimodal Emotion Recognition}.
\newblock \bibinfo{journal}{\emph{IEEE Access}}  \bibinfo{volume}{7}
  (\bibinfo{year}{2019}), \bibinfo{pages}{172948--172964}.
\newblock
\urldef\tempurl%
\url{https://doi.org/10.1109/ACCESS.2019.2955637}
\showDOI{\tempurl}


\bibitem[Nottage and Horder(2016)]%
        {high}
\bibfield{author}{\bibinfo{person}{Judith~F Nottage} {and}
  \bibinfo{person}{Jamie Horder}.} \bibinfo{year}{2016}\natexlab{}.
\newblock \showarticletitle{State-of-the-art analysis of high-frequency (gamma
  range) electroencephalography in humans}.
\newblock \bibinfo{journal}{\emph{Neuropsychobiology}} \bibinfo{volume}{72},
  \bibinfo{number}{3-4} (\bibinfo{year}{2016}), \bibinfo{pages}{219--228}.
\newblock
\urldef\tempurl%
\url{https://doi.org/10.1159/000382023}
\showDOI{\tempurl}


\bibitem[Olsen(2012)]%
        {olsen2012tobii}
\bibfield{author}{\bibinfo{person}{Anneli Olsen}.}
  \bibinfo{year}{2012}\natexlab{}.
\newblock \showarticletitle{The Tobii IVT Fixation Filter Algorithm
  description}.
\newblock
\urldef\tempurl%
\url{https://api.semanticscholar.org/CorpusID:52834703}
\showURL{%
\tempurl}


\bibitem[Park et~al\mbox{.}(2014)]%
        {park2014human}
\bibfield{author}{\bibinfo{person}{Ukeob Park}, \bibinfo{person}{Rammohan
  Mallipeddi}, {and} \bibinfo{person}{Minho Lee}.}
  \bibinfo{year}{2014}\natexlab{}.
\newblock \showarticletitle{Human implicit intent discrimination using EEG and
  eye movement}. In \bibinfo{booktitle}{\emph{International Conference on
  Neural Information Processing}}. Springer, \bibinfo{pages}{11--18}.
\newblock
\urldef\tempurl%
\url{https://doi.org/10.1007/978-3-319-12637-1_2}
\showDOI{\tempurl}


\bibitem[Roy et~al\mbox{.}(2017)]%
        {roy2017gaussian}
\bibfield{author}{\bibinfo{person}{Vandana Roy}, \bibinfo{person}{Shailja
  Shukla}, \bibinfo{person}{Piyush~Kumar Shukla}, {and} \bibinfo{person}{Paresh
  Rawat}.} \bibinfo{year}{2017}\natexlab{}.
\newblock \showarticletitle{Gaussian elimination-based novel canonical
  correlation analysis method for EEG motion artifact removal}.
\newblock \bibinfo{journal}{\emph{Journal of Healthcare Engineering}}
  \bibinfo{volume}{2017} (\bibinfo{year}{2017}).
\newblock
\urldef\tempurl%
\url{https://doi.org/10.1155/2017/9674712}
\showDOI{\tempurl}


\bibitem[Sattar et~al\mbox{.}(2015)]%
        {sattar2015prediction}
\bibfield{author}{\bibinfo{person}{Hosnieh Sattar}, \bibinfo{person}{Sabine
  Müller}, \bibinfo{person}{Mario Fritz}, {and} \bibinfo{person}{Andreas
  Bulling}.} \bibinfo{year}{2015}\natexlab{}.
\newblock \showarticletitle{Prediction of search targets from fixations in
  open-world settings}. In \bibinfo{booktitle}{\emph{2015 IEEE Conference on
  Computer Vision and Pattern Recognition (CVPR)}}. \bibinfo{pages}{981--990}.
\newblock
\urldef\tempurl%
\url{https://doi.org/10.1109/CVPR.2015.7298700}
\showDOI{\tempurl}


\bibitem[Sharma et~al\mbox{.}(2023)]%
        {sharma2023towards}
\bibfield{author}{\bibinfo{person}{Mansi Sharma}, \bibinfo{person}{Maurice
  Rekrut}, \bibinfo{person}{Jan Alexandersson}, {and} \bibinfo{person}{Antonio
  Kr{\"u}ger}.} \bibinfo{year}{2023}\natexlab{}.
\newblock \showarticletitle{Towards Improving EEG-Based Intent Recognition in
  Visual Search Tasks}. In \bibinfo{booktitle}{\emph{Neural Information
  Processing: 29th International Conference, ICONIP, Proceedings, Part III}}.
  Springer, \bibinfo{pages}{604--615}.
\newblock
\urldef\tempurl%
\url{https://doi.org/10.1007/978-3-031-30111-7_51}
\showDOI{\tempurl}


\bibitem[Slanzi et~al\mbox{.}(2017)]%
        {slanzi2017combining}
\bibfield{author}{\bibinfo{person}{Gino Slanzi}, \bibinfo{person}{Jorge~A.
  Balazs}, {and} \bibinfo{person}{Juan~D. Velásquez}.}
  \bibinfo{year}{2017}\natexlab{}.
\newblock \showarticletitle{Combining eye tracking, pupil dilation and EEG
  analysis for predicting web users click intention}.
\newblock \bibinfo{journal}{\emph{Information Fusion}}  \bibinfo{volume}{35}
  (\bibinfo{year}{2017}), \bibinfo{pages}{51--57}.
\newblock
\showISSN{1566-2535}
\urldef\tempurl%
\url{https://doi.org/10.1016/j.inffus.2016.09.003}
\showDOI{\tempurl}


\bibitem[Stauden et~al\mbox{.}(2018)]%
        {stauden2018visual}
\bibfield{author}{\bibinfo{person}{Sven Stauden}, \bibinfo{person}{Michael
  Barz}, {and} \bibinfo{person}{Daniel Sonntag}.}
  \bibinfo{year}{2018}\natexlab{}.
\newblock \bibinfo{booktitle}{\emph{Visual Search Target Inference Using Bag of
  Deep Visual Words: 41st German Conference on AI, 2018, Proceedings}}.
\newblock \bibinfo{pages}{297--304}.
\newblock
\urldef\tempurl%
\url{https://doi.org/10.1007/978-3-030-00111-7_25}
\showDOI{\tempurl}


\bibitem[Steil et~al\mbox{.}(2018)]%
        {steil18_mobilehci}
\bibfield{author}{\bibinfo{person}{Julian Steil}, \bibinfo{person}{Philipp
  M{\"{u}}ller}, \bibinfo{person}{Yusuke Sugano}, {and}
  \bibinfo{person}{Andreas Bulling}.} \bibinfo{year}{2018}\natexlab{}.
\newblock \showarticletitle{Forecasting User Attention During Everyday Mobile
  Interactions Using Device-Integrated and Wearable Sensors}. In
  \bibinfo{booktitle}{\emph{Proceedings of ACM International Conference on
  Human-Computer Interaction with Mobile Devices and Services (MobileHCI)}}.
  \bibinfo{pages}{1--13}.
\newblock
\urldef\tempurl%
\url{https://doi.org/10.1145/3229434.3229439}
\showDOI{\tempurl}


\bibitem[Strohm et~al\mbox{.}(2021)]%
        {strohm21_iccv}
\bibfield{author}{\bibinfo{person}{Florian Strohm}, \bibinfo{person}{Ekta
  Sood}, \bibinfo{person}{Sven Mayer}, \bibinfo{person}{Philipp Müller},
  \bibinfo{person}{Mihai Bâce}, {and} \bibinfo{person}{Andreas Bulling}.}
  \bibinfo{year}{2021}\natexlab{}.
\newblock \showarticletitle{Neural Photofit: Gaze-based Mental Image
  Reconstruction}. In \bibinfo{booktitle}{\emph{Proceedings of IEEE
  International Conference on Computer Vision (ICCV)}}.
  \bibinfo{pages}{245--254}.
\newblock
\urldef\tempurl%
\url{https://doi.org/10.1109/ICCV48922.2021.00031}
\showDOI{\tempurl}


\bibitem[Trabulsi et~al\mbox{.}(2021)]%
        {trabulsi2021optimizing}
\bibfield{author}{\bibinfo{person}{Julia Trabulsi}, \bibinfo{person}{Kian
  Norouzi}, \bibinfo{person}{Seidi Suurmets}, \bibinfo{person}{Mike Storm},
  {and} \bibinfo{person}{Thomas~Zo{\"e}ga Rams{\o}y}.}
  \bibinfo{year}{2021}\natexlab{}.
\newblock \showarticletitle{Optimizing fixation filters for eye-tracking on
  small screens}.
\newblock \bibinfo{journal}{\emph{Frontiers in neuroscience}}
  (\bibinfo{year}{2021}), \bibinfo{pages}{1257}.
\newblock
\urldef\tempurl%
\url{https://doi.org/10.3389/fnins.2021.578439}
\showDOI{\tempurl}


\bibitem[Ward et~al\mbox{.}(2016)]%
        {ward2016possibility}
\bibfield{author}{\bibinfo{person}{Nigel~G. Ward}, \bibinfo{person}{Chelsey~N.
  Jurado}, \bibinfo{person}{Ricardo~A. Garcia}, {and}
  \bibinfo{person}{Florencia~A. Ramos}.} \bibinfo{year}{2016}\natexlab{}.
\newblock \showarticletitle{On the Possibility of Predicting Gaze Aversion to
  Improve Video-Chat Efficiency}. In \bibinfo{booktitle}{\emph{Proceedings of
  the Ninth Biennial ACM Symposium on Eye Tracking Research \& Applications}}.
  \bibinfo{pages}{267–270}.
\newblock
\urldef\tempurl%
\url{https://doi.org/10.1145/2857491.2857497}
\showDOI{\tempurl}


\bibitem[Zander et~al\mbox{.}(2014)]%
        {zander2014towards}
\bibfield{author}{\bibinfo{person}{Thorsten Zander}, \bibinfo{person}{Jonas
  Brönstrup}, \bibinfo{person}{Romy Lorenz}, {and} \bibinfo{person}{Laurens
  Krol}.} \bibinfo{year}{2014}\natexlab{}.
\newblock \bibinfo{booktitle}{\emph{Towards BCI-Based Implicit Control in
  Human–Computer Interaction}}.
\newblock \bibinfo{pages}{67--90}.
\newblock
\urldef\tempurl%
\url{https://doi.org/10.1007/978-1-4471-6392-3_4}
\showDOI{\tempurl}


\bibitem[Zhao et~al\mbox{.}(2020)]%
        {zhao2020research}
\bibfield{author}{\bibinfo{person}{Minrui Zhao}, \bibinfo{person}{Hongni Gao},
  \bibinfo{person}{Wei Wang}, {and} \bibinfo{person}{Jue Qu}.}
  \bibinfo{year}{2020}\natexlab{}.
\newblock \showarticletitle{Research on Human-Computer Interaction Intention
  Recognition Based on EEG and Eye Movement}.
\newblock \bibinfo{journal}{\emph{IEEE Access}}  \bibinfo{volume}{8}
  (\bibinfo{year}{2020}), \bibinfo{pages}{145824--145832}.
\newblock
\urldef\tempurl%
\url{https://doi.org/10.1109/ACCESS.2020.3011740}
\showDOI{\tempurl}


\end{thebibliography}

% \appendix









\end{document}